\pdfoutput=1

\documentclass[a4paper,11pt]{article}
\usepackage{jheppub}

\usepackage{bm}

\usepackage{color}
\usepackage[usenames,dvipsnames,svgnames,table]{xcolor}

\usepackage{amsmath}
\usepackage{amssymb}
\usepackage{graphicx}
\usepackage{slashed}
\usepackage{soul}
\usepackage{lscape}
\usepackage{graphicx}
\usepackage{xcolor}
\usepackage{multirow}
\usepackage{placeins}

\newcommand{\ccGyp}{Ca(SO$_4)\!\cdot\!2$(H$_2$O)}
\newcommand{\ccHal}{NaCl}

\newcommand{\ccSin}{CaCl$_2 \! \cdot\! 2($H$_2$O)}

\newcommand{\ccNch}{Mn$^{2+}_2$SiO$_3$(OH)$_2 \! \cdot \! ($H$_2$O)}

\newcommand{\ccOli}{Mg$_{1.6}$Fe$^{2+}_{0.4}$(SiO$_4$)}
\newcommand{\ccPhl}{KMg$_3$AlSi$_3$O$_{10}$F(OH)}

\newcommand{\Ur}{\ensuremath{ { ^{238}{\rm U} } } }
\newcommand{\Th}{\ensuremath{ { ^{234}{\rm Th} } } }

\newcommand{\linkPaSpec}{https://github.com/sbaum90/paleoSpec}
\newcommand{\linkPaSens}{https://github.com/sbaum90/paleoSens}

\begin{document}

\title{New Projections for Dark Matter Searches with Paleo-Detectors}

\author[a]{Sebastian~Baum,}
\author[b]{Thomas~D.~P.~Edwards,}
\author[c,d,b]{Katherine~Freese,}
\author[e,f,g]{Patrick~Stengel}

\affiliation[a]{Stanford Institute for Theoretical Physics, Stanford University, Stanford, CA~94305, USA}
\affiliation[b]{The Oskar Klein Centre for Cosmoparticle Physics, Department of Physics, Stockholm University, Alba Nova, 10691~Stockholm, Sweden}
\affiliation[c]{Department of Physics, University of Texas, Austin, TX 78722, USA}
\affiliation[d]{Nordita, KTH Royal Institute of Technology and Stockholm University, Roslagstullsbacken~23, 10691~Stockholm, Sweden}
\affiliation[e]{Scuola Internazionale Superiore di Studi Avanzati (SISSA), via Bonomea 265, 34136 Trieste, Italy}
\affiliation[f]{INFN, Sezione di Trieste, via Valerio 2, 34127 Trieste, Italy}
\affiliation[g]{Institute for Fundamental Physics of the Universe (IFPU), via Beirut 2, 34151 Trieste, Italy}

\emailAdd{sbaum@stanford.edu}
\emailAdd{thomas.edwards@fysik.su.se}
\emailAdd{ktfreese@utexas.edu}
\emailAdd{pstengel@sissa.it}

\abstract{
Paleo-detectors are a proposed experimental technique to search for dark matter (DM). In lieu of the conventional approach of operating a tonne-scale real-time detector to search for DM-induced nuclear recoils, paleo-detectors take advantage of small samples of naturally occurring rocks on Earth that have been deep underground ($\gtrsim 5\,$km), accumulating nuclear damage tracks from recoiling nuclei for $\mathcal{O}(1)\,$Gyr. Modern microscopy techniques promise the capability to read out nuclear damage tracks with nanometer resolution in macroscopic samples. Thanks to their $\mathcal{O}(1)\,$Gyr integration times, paleo-detectors could constitute nuclear recoil detectors with keV recoil energy thresholds and 100\,kilotonne-yr exposures. This combination would allow paleo-detectors to probe DM-nucleon cross sections orders of magnitude below existing upper limits from conventional direct detection experiments. In this article, we use improved background modeling and a new spectral analysis technique to update the sensitivity forecast for paleo-detectors. We demonstrate the robustness of the sensitivity forecast to the (lack of) ancillary measurements of the age of the samples and the parameters controlling the backgrounds, systematic mismodeling of the spectral shape of the backgrounds, and the radiopurity of the mineral samples. Specifically, we demonstrate that even if the uranium concentration in paleo-detector samples is $10^{-8}$ (per weight), many orders of magnitude larger than what we expect in the most radiopure samples obtained from ultra basic rock or marine evaporite deposits, paleo-detectors could still probe DM-nucleon cross sections below current limits. For DM masses $\lesssim 10\,$GeV/$c^2$, the sensitivity of paleo-detectors could still reach down all the way to the conventional neutrino floor in a Xe-based direct detection experiment.
}

\maketitle
\flushbottom

\section{Introduction}

More than a century after evidence for the existence of Dark Matter (DM) in our Universe started to emerge~\cite{Bertone:2016nfn}, we still do not know what DM is made out of. Weakly Interacting Massive Particles (WIMPs) remain one of the best-motivated DM candidates to date~\cite{Jungman:1995df,Arcadi:2017kky,Roszkowski:2017nbc}. Among the most sensitive probes of the WIMP DM paradigm are {\it direct detection} experiments~\cite{Drukier:1983gj,Goodman:1984dc,Drukier:1986tm,Spergel:1987kx,Collar:1992qc}, where one searches for interactions of DM with atomic nuclei. Over the past few decades, direct detection experiments have made impressive progress: liquid noble gas detectors have reached $\mathcal{O}(\mbox{tonne-yr})$ exposures with nuclear recoil energy thresholds as small as a few keV~\cite{Akerib:2016vxi,Agnes:2018fwg,Aprile:2018dbl,Wang:2020coa}, while cryogenic bolometric detectors have pushed the nuclear recoil energy threshold below 100\,eV, albeit as yet only with detectors accumulating $\mathcal{O}(\mbox{kg\,days})$ of exposure~\cite{Angloher:2015ewa,Armengaud:2016cvl,Agnese:2017jvy,Petricca:2017zdp,Armengaud:2019kfj}. A large experimental program is underway to extend the sensitivity of direct detection experiments both to lower WIMP masses and smaller WIMP-nucleon cross sections~\cite{Aprile:2015uzo,Mount:2017qzi,Aalbers:2016jon,Aalseth:2017fik,Amaudruz:2017ibl}, as well as to develop directional detectors~\cite{Daw:2011wq,Cappella:2013rua,Battat:2014van,Riffard:2013psa,Santos:2013hpa,Monroe:2012qma,Leyton:2016nit,Miuchi:2010hn,Nakamura:2015iza, Battat:2016xxe, Vahsen:2020pzb}. However, these detectors are becoming ever more costly and more challenging to operate.

{\it Paleo-detectors}~\cite{Baum:2018tfw,Drukier:2018pdy,Edwards:2018hcf} are a proposed alternative technique to search for interactions of DM with nuclei: instead of operating a nuclear recoil detector in real time, as in a conventional direct detection experiment, one would search for the persistent traces of WIMP-nucleon interactions recorded in natural minerals over timescales as long as $\mathcal{O}(1)\,$Gyr. Many minerals are excellent solid state nuclear track detectors~\cite{Fleischer:1964,Fleischer383,Fleischer:1965yv,GUO2012233} and retain these traces (i.e. damage tracks) caused by a recoiling atomic nucleus for timescales which can exceed the age of the Earth by many orders of magnitude (i.e. $\gg 4.5\,\mathrm{Gyr}$). Natural minerals commonly found on Earth are up to $\mathcal{O}(1)\,$Gyr old and can be dated to a few-percent accuracy~\cite{GTS2012,Gallagher:1998,vandenHaute:1998}. Modern microscopy technologies, including electron microscopy~\cite{Fleischer:1965yv, Toulemonde:2006}, Helium Ion Beam Microscopy (HIBM)~\cite{HILL201265, VANGASTEL20122104}, Small Angle X-ray scattering (SAXs)~\cite{RODRIGUEZ2014150,SAXS3d,SAXSres}, and optical microscopy~\cite{GUO2012233,BARTZ2013273,Kouwenberg:2018} potentially allow for the readout of such nuclear damage tracks in macroscopic samples of target material. For example, HIBM may allow one to read out $\mathcal{O}(10)\,$mg of material with $\mathcal{O}(1)\,$nm precision when combined with pulsed-laser and fast-ion-beam ablation techniques~\cite{bioHIB,ECHLIN20151,PFEIFENBERGER2017109,Randolph:2018}. On the other hand, SAXs promises spatial resolutions of $\mathcal{O}(15)\,$nm in $\mathcal{O}(100)\,$g of target material~\cite{SAXS3d,SAXSres} (see the discussion in Refs.~\cite{Baum:2018tfw,Drukier:2018pdy}). These advanced microscopy techniques enable paleo-detectors to reach much larger exposures with better track length resolutions than previous ideas to probe rare events using natural minerals; see Refs.~\cite{Goto:1958,Goto:1963zz,Fleischer:1969mj,Fleischer:1970zy,Fleischer:1970vm,Alvarez:1970zu,Kolm:1971xb,Eberhard:1971re,Ross:1973it,Price:1983ax,Kovalik:1986zz,Price:1986ky,Ghosh:1990ki,Jeon:1995rf,Collar:1999md} and, in particular, the work by Snowden-Ifft {\it et al.} searching for damage tracks from WIMP-nucleus interactions in ancient phlogopite mica samples~\cite{SnowdenIfft:1995ke,Collar:1994mj,Engel:1995gw,SnowdenIfft:1997hd}. While not the focus of this article, paleo-detectors have also been proposed as a probe of astrophysical neutrino fluxes from Galactic supernovae~\cite{Baum:2019fqm}, cosmic ray interactions in Earth's atmosphere~\cite{Jordan:2020gxx}, and from our Sun~\cite{Arellano:2021jul}. There has been recent related work on using crystal-defects created in certain minerals by WIMP-nucleus interactions to search for light DM~\cite{Rajendran:2017ynw,Essig:2016crl,Budnik:2017sbu} or to monitor nuclear reactors~\cite{Cogswell:2021qlq}, and to use very long damage tracks in ancient quartz samples to search for ultra-heavy DM~\cite{Ebadi:2021cte}.

Using natural minerals found on Earth to record DM-nucleus interactions over $\mathcal{O}(100)\,{\rm Myr} - \mathcal{O}(1)\,$Gyr timescales provides paleo-detectors with two crucial advantages over conventional direct detection experiments. First, the enormous integration times promise exposures many orders of magnitude larger than what is feasible in laboratory experiments. Reading out 100\,g of material that has been recording nuclear damage tracks for 1\,Gyr yields an exposure of $\varepsilon = 100\,$g\,Gyr, or, in units more appropriate for a real-time experiment, $\varepsilon = 100\,$kilotonne\,yr. The $\mathcal{O}(15)\,$nm spatial resolution achievable with SAXs readout corresponds to a nuclear recoil energy threshold of $\mathcal{O}(1)\,$keV. Thus, paleo-detectors would combine the recoil energy threshold achievable in direct detection experiments utilizing liquid noble gases as target materials with the exposures achievable in large (planned) neutrino detectors such as Super-/Hyper-Kamiokande or DUNE. Second, while conventional direct detection experiments measure events in real time, paleo-detectors integrate over timescales up to $\mathcal{O}(1)\,$Gyr, comparable to the age of the Solar System, $t_\odot \sim 4.5\,$Gyr, and the orbital period of the Solar System around the Milky Way, $T_\odot \sim 250\,$Myr. Using a series of samples with different ages, paleo-detectors offer a unique pathway to gaining information about a signal's time-evolution over Gyr timescales~\cite{Baum:2019fqm,Jordan:2020gxx,Arellano:2021jul,SUBSTRUCTURE}. 

Of course, paleo-detectors must pay a price for these advantages: compared to a conventional detector operated in a controlled laboratory environment, background mitigation might be much more challenging. The background sources for DM searches in paleo-detectors are qualitatively the same as in conventional direct detection experiments since both approaches are, at their heart, nuclear recoil searches. Extensive background studies for paleo-detectors have been performed in previous work~\cite{Baum:2018tfw,Drukier:2018pdy,Baum:2019fqm} and we will discuss the different sources further below. As we will see, the most important background sources are cosmic rays, solar neutrinos, and radioactivity. Regarding the first, the fundamental strategy for paleo-detectors is to use natural minerals extracted from deep within the Earth where the rock-overburden has shielded them from cosmogenic backgrounds. Conventional direct detection experiments are operated in deep underground laboratories for the same reason. However, paleo-detectors require only (at most) a few kg of target material compared to actively operating a tonne-scale experiment; samples of such modest sizes can be extracted from much greater depths than those of existing underground laboratories (for example from existing boreholes drilled for geological R\&D and oil exploration). For the backgrounds from solar neutrinos and radioactivity it is important to note that, due to their large exposures, paleo-detectors would contain a sizable number of background events. A rare event search, such as for DM-induced nuclear recoils, would thus not proceed as in conventional direct detection experiments where one typically attempts to find a background-free signal region. Instead, a paleo-detector analysis is more similar to a collider physics search: one would model the distribution of background events (for example in track-length space) and search for deviations of the observed data from the background-only prediction using a signal template. One important consequence of the large number of background (and possible signal) events is that one can characterize many aspects of the backgrounds from the data themselves. Just as in collider searches, paleo-detectors can use {\it control regions}, where the backgrounds dominate, to better characterize the background contribution in the {\it signal region} of the search.

The remainder of this article is organized as follows: in Section~\ref{sec:DMsig} we discuss the calculation of DM signals in paleo-detectors followed by the most important background sources in Section~\ref{sec:Bkg}. In Section~\ref{sec:Sens} we show projections for the reach of paleo-detectors in the conventional WIMP mass--cross section plane. Compared to previous estimates of the sensitivity of paleo-detectors to DM signals~\cite{Baum:2018tfw,Drukier:2018pdy,Edwards:2018hcf}, in this work we make use of both improved background modeling and new statistical techniques. In this sense, the results presented here supersede the sensitivity projections shown in Refs.~\cite{Baum:2018tfw,Drukier:2018pdy,Edwards:2018hcf}. In Section~\ref{sec:Sens} we also explore the robustness of the sensitivity projections to changes in the assumptions we make in our analysis; for example, we remove external constraints on the age of the minerals and the parameters controlling the background normalizations, and explore the effect of systematic mismodeling of the spectral shape of the backgrounds. Furthermore, we show how the projected sensitivity changes with the radiopurity of the samples, and demonstrate that, even if the uranium concentration in paleo-detector samples is $10^{-8}$ (per weight, denoted as ``g/g'' throughout), many orders of magnitude larger than what we expect in the most radiopure samples obtained from ultra basic rock or marine evaporite deposits, paleo-detectors could still probe DM-nucleon cross sections below current limits. Specifically, for WIMP masses $m_\chi \lesssim 10\,$GeV/$c^2$, the sensitivity of paleo-detectors could still reach down all the way to the conventional neutrino floor in a Xe-based direct detection experiment. We summarize and conclude in Section~\ref{sec:Conc}. We make the code used in this work available: \href{\linkPaSpec}{\texttt{paleoSpec}}~\cite{PaleoSpec} for the computation of the signal and background spectra, and \href{\linkPaSens}{\texttt{paleoSens}}~\cite{PaleoSens} for the sensitivity forecasts.

\section{Dark Matter Signals in Paleo-Detectors} \label{sec:DMsig}

The signature of WIMP DM in a paleo-detector is the spectrum of damage tracks left by nuclei in the mineral sample after receiving a ``kick'' from WIMP-nucleus scattering. In this section, we discuss how this signal is computed. The differential rate of nuclear recoils per unit target mass for a WIMP with mass $m_\chi$ scattering off target nuclei $N$ with recoil energy $E_R$ is given by~\cite{Engel:1991wq,Engel:1992bf}

\begin{equation} \label{eq:recoilSpec}
   \left( \frac{d R}{d E_R} \right)_N = \frac{A_N^2 \sigma_p^{\rm SI} F_N^2(E_R)}{2 \mu_p^2 m_\chi} \rho_\chi \eta_\chi(v_{\rm min}) \;,
\end{equation}
where $A_N$ is the atomic mass number of $N$, $\mu_p = m_p m_\chi/(m_p + m_\chi)$is the reduced mass of the WIMP-proton system with the proton mass $m_p$, and we have assumed isospin-conserving spin-independent WIMP-nucleon interactions with the (zero momentum transfer) WIMP-proton cross section $\sigma_p^{\rm SI}$. Expressions analogous to Eq.~\eqref{eq:recoilSpec} for isospin-violating SI interaction and spin-dependent WIMP-nucleon interactions can be found in Refs.~\cite{Engel:1991wq,Engel:1992bf,Ressell:1993qm,Bednyakov:2004xq,Bednyakov:2006ux}, and for more general WIMP-nucleus interactions in Refs.~\cite{Fan:2010gt,Fitzpatrick:2012ix} . Note that while we focus on spin-independent interactions in this work, paleo-detectors could also probe spin-dependent WIMP-nucleus interactions, see Ref.~\cite{Drukier:2018pdy}. 

In Eq.~\eqref{eq:recoilSpec}, the nuclear form factor $F_N(E_R)$ accounts for the finite size of the nucleus; we will use the Helm parameterization~\cite{Helm:1956zz,Lewin:1995rx,Duda:2006uk} in our numerical calculations.\footnote{Note that more refined calculations of the form factors are available, although only for a few isotopes, see e.g. Refs.~\cite{Vietze:2014vsa,Gazda:2016mrp,Korber:2017ery,Hoferichter:2018acd}.} The distribution of WIMPs is described by the local WIMP (mass) density, $\rho_\chi$, and the {\it mean inverse speed},

\begin{equation} \label{eq:eta}
   \eta_\chi(v_{\rm min}) = \int_{v \geq v_{\rm min}} d^3{\bm v}\;\frac{f({\bm v})}{v} \;,
\end{equation}
where $f({\bm v})$ is the velocity distribution of WIMPs with ${\bm v}$ ($v = \left|{\bm v}\right|$) the velocity (speed) of the WIMP relative to the detector. In Eq.~\eqref{eq:eta}, $v_{\rm min} = \sqrt{m_N E_R / 2 \mu_N^2}$ accounts for the scattering kinematics; $v_{\rm min}$ is the minimal speed, with respect to a nucleus of mass $m_N$, that a WIMP must have to give rise to a recoil with energy $E_R$ (where $\mu_N$ is the reduced mass of the WIMP-nucleus system). In our numerical calculations we set $\rho_\chi = 0.3\,$GeV/cm$^3$ while for $f({\bm v})$ we use a Maxwell-Boltzmann velocity distribution in the Galactic rest frame with velocity dispersion $\sigma_v = 166\,$km/s~\cite{Koposov:2009hn}, truncated at the escape speed $v_{\rm esc} = 550\,$km/s~\cite{Piffl:2013mla}, and boosted to the Solar System frame by $v_\odot = 248\,$km/s~\cite{Bovy:2012ba} as in the Standard Halo Model~(SHM)~\cite{Drukier:1986tm,Lewin:1995rx,Freese:2012xd}. 

\begin{figure}
   \includegraphics[width=0.49\linewidth]{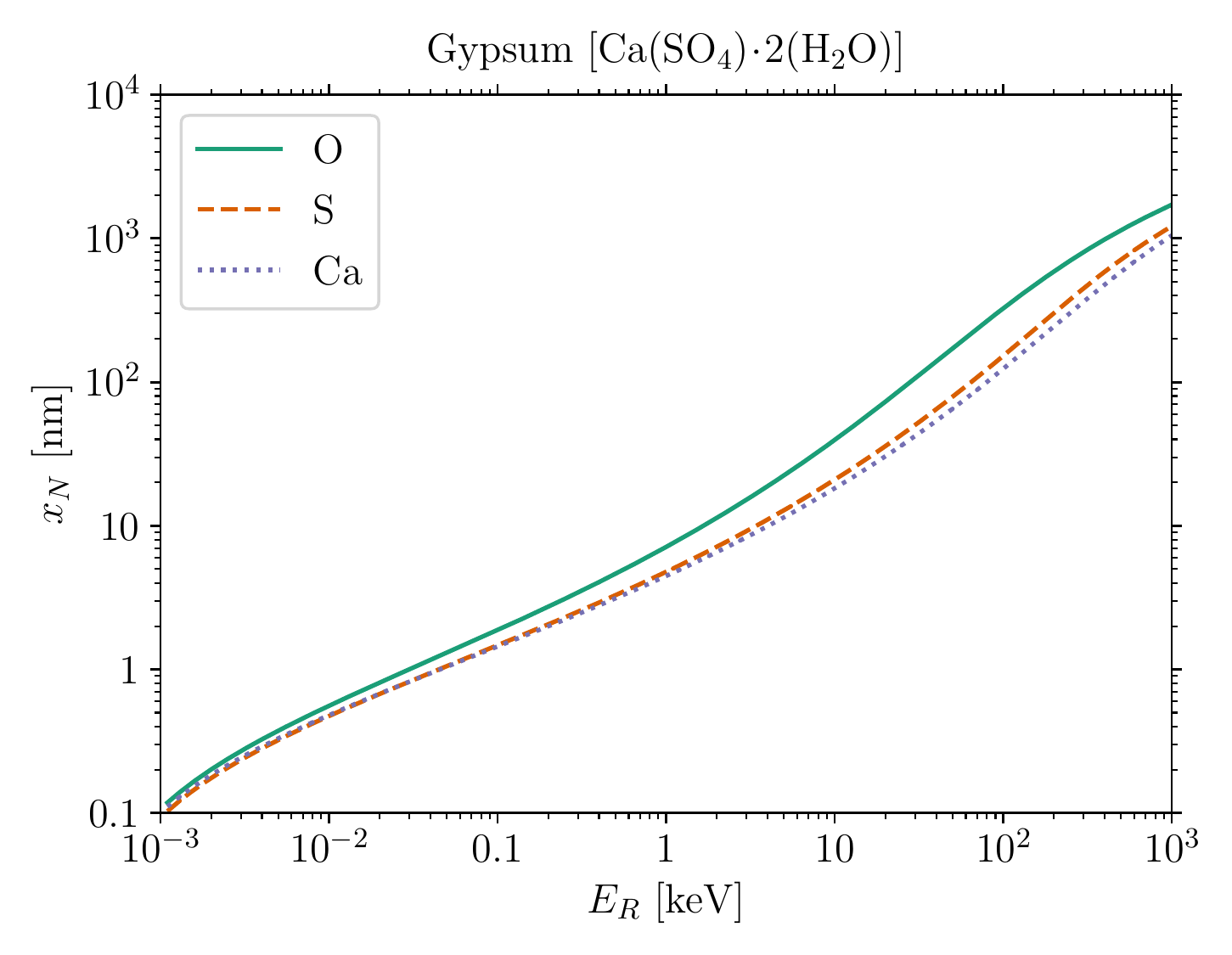}
   \includegraphics[width=0.49\linewidth]{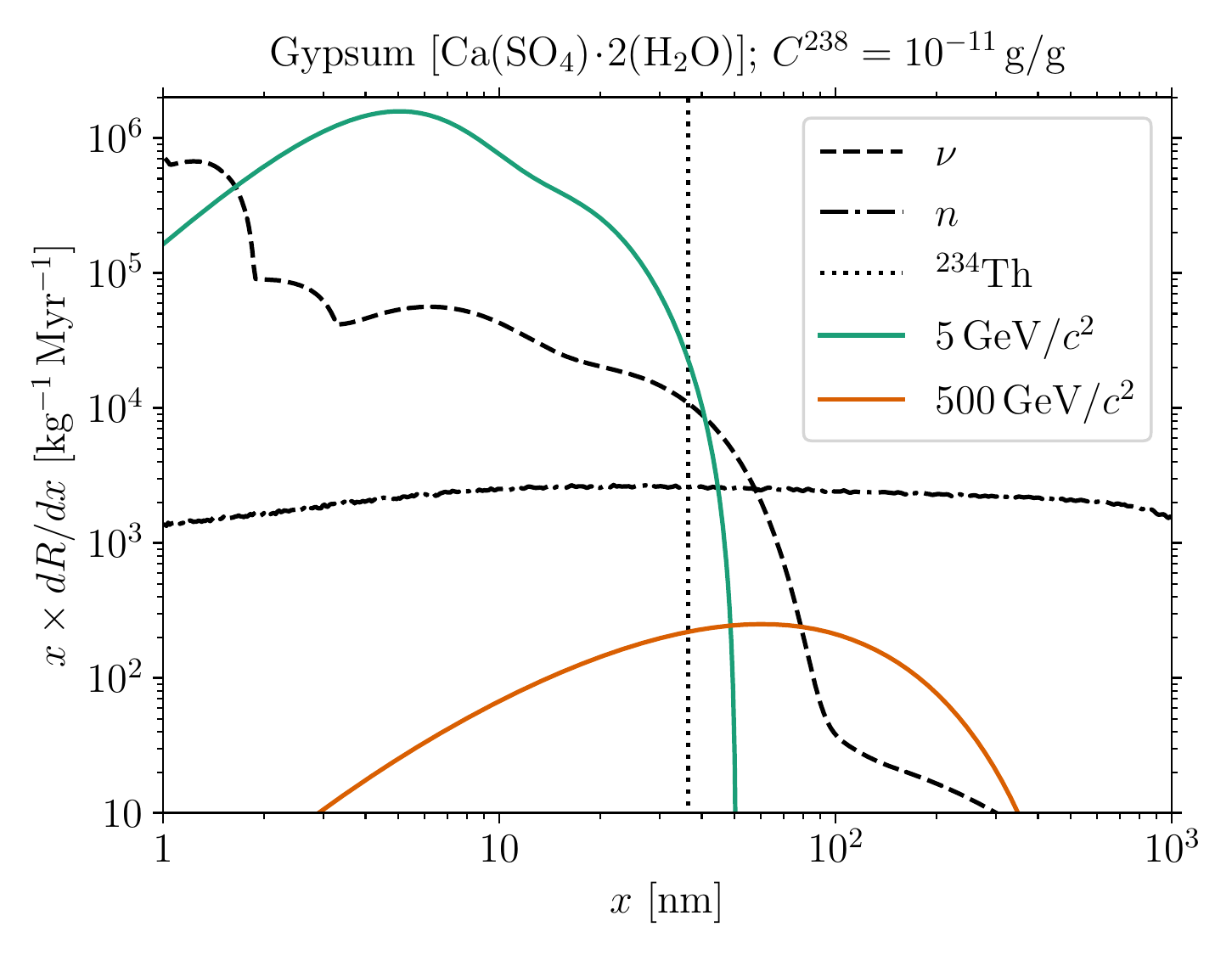}
   \caption{{\bf Left:}~Range $x_N$ of different ions with initial energy $E_R$ in gypsum [\ccGyp]. {\bf Right:}~Differential rate of tracks per track length and unit target mass in gypsum from a $5\,$GeV/$c^2$ and a $500\,$GeV/$c^2$ WIMP compared to background spectra induced by neutrinos ($\nu$), radiogenic neutrons ($n$), and $\Ur \to \Th + \alpha$ recoils (\Th), see the discussion in Section~\ref{sec:Bkg}. For the normalization of the WIMP DM spectra we set the spin-independent WIMP-nucleon cross sections to $\sigma_p^{\rm SI} = 10^{-43}\,{\rm cm}^2$ for the 5\,GeV/$c^2$ case and $\sigma_p^{\rm SI} = 10^{-46}\,{\rm cm}^2$ for the 500\,GeV/$c^2$ case, compatible with current upper bounds~\cite{Aprile:2018dbl,Aprile:2019xxb}.}
   \label{fig:RangeSpec}
\end{figure}

In order to obtain the differential track length spectrum (per unit target mass), $dR/dx$, we use the range of a nucleus $N$ with energy $E_R$,

\begin{equation} \label{eq:range}
   x_N(E_R) = \int_0^{E_R} dE \; \left| \frac{dE}{dx} \right|_N^{-1} \;,
\end{equation}
where $\left(dE/dx\right)_N$ is the stopping power of $N$ in the target material. The true length of a damage track in a given target may differ from the range if, for example, the nucleus creates a lasting damage track only along some portion of the length it travels through the material, or if the nucleus' trajectory deviates significantly from a straight line. Our previous studies suggest that such effects should be small (see the discussion in Ref.~\cite{Drukier:2018pdy}), making Eq.~\eqref{eq:range} a good proxy for the track length. Furthermore, the effects of thermal annealing could potentially be significant over geological timescales; fortunately, any associated modifications to the track lengths would be similar for both the signal and background recoils. While detailed experimental studies for each target material will be required to account for these possible corrections, for the purposes of this work we will use Eq.~\eqref{eq:range} for the track length and \texttt{SRIM}~\cite{Ziegler:1985,Ziegler:2010} to calculate stopping powers. The left panel of Figure~\ref{fig:RangeSpec} shows the range of the different constituent nuclei\footnote{Note that we do not show the range of H in Figure~\ref{fig:RangeSpec} since we do not expect ions with charge $Z \leq 2$ (in particular, H and He) to produce lasting damage tracks in most materials, see the discussion in~\cite{Drukier:2018pdy}.} in gypsum~[\ccGyp] as an example; we see that a 1\,keV recoil gives rise to a few-nm long track, while a 100\,keV recoil produces a track with a length of a few hundred nm.

In a composite target material composed of different isotopes with mass fraction $\xi_N$, the differential rate at which tracks are produced (per unit target mass) is then given by

\begin{equation} \label{eq:dRdx}
   \frac{dR}{dx} = \sum_N \xi_N \left(\frac{dR}{dE_R} \right)_N \left(\frac{dE}{d x} \right)_N \;,
\end{equation}
with $dR/dE_R$ given in Eq.~\eqref{eq:recoilSpec}. In the right panel of Figure~\ref{fig:RangeSpec} we show $dR/dx$ for a 5\,GeV/$c^2$ (with cross section $\sigma_p^{\rm SI} = 10^{-43}\,{\rm cm}^2$) and a 500\,GeV/$c^2$ (with $\sigma_p^{\rm SI} = 10^{-46}\,{\rm cm}^2$) WIMP in gypsum as an example together with various background spectra which we will discuss in Section~\ref{sec:Bkg}. Note that for the $5\,$GeV/$c^2$ case, the largest contribution to the signal is at track lengths of a few nm, while heavier WIMPs, such as the 500\,GeV/$c^2$ case shown in Figure~\ref{fig:RangeSpec}, give rise to longer tracks with typical lengths of order $100\,$nm. 

The nuclear recoil tracks in a paleo-detector sample can be read out with a variety of readout techniques as briefly discussed in the introduction (see Ref.~\cite{Drukier:2018pdy} for a more detailed discussion). Here, we will consider two scenarios: 
\begin{itemize}
   \item {\it High-resolution readout scenario:} We assume that 10\,mg of material can be read out with a track length resolution of $\sigma_x = 1\,$nm. This scenario may be achievable with Helium Ion Beam Microscopy (HIBM) combined with pulsed-laser and fast-ion-beam ablation techniques~\cite{HILL201265, VANGASTEL20122104,bioHIB,ECHLIN20151,PFEIFENBERGER2017109,Randolph:2018}. As we will see, this scenario is advantageous for low-mass ($m_\chi \lesssim 10\,$GeV$/c^2$) WIMP searches.
   \item {\it High-exposure readout scenario:} We assume that 100\,g of material can be read out with $\sigma_x = 15\,$nm. This scenario could be realized via Small Angle X-ray scattering (SAXs) tomography at a synchrotron facility~\cite{RODRIGUEZ2014150,SAXS3d,SAXSres}; it is better suited for heavier ($m_\chi \gtrsim 10\,$GeV$/c^2$) WIMP searches.
\end{itemize}
These are the same readout scenarios as what we have used in previous paleo-detector studies~\cite{Baum:2018tfw,Drukier:2018pdy,Edwards:2018hcf,Baum:2019fqm,Jordan:2020gxx}. We note that these scenarios represent challenging applications of the respective techniques; demonstrating their feasibility and especially their scalability to the proposed sample masses would be an important step towards the realization of a paleo-detector experiment. There are ongoing efforts to demonstrate the readout of few-keV recoils using a variety of microscopy techniques at JAMSTEC~\cite{Hirose} and at Rensselaer Polytechnic Institute~\cite{RPI}.

In order to assess the sensitivity of paleo-detectors, we will consider the binned track length spectrum as the experimental observable. Accounting for the finite resolution of the readout method, the number of tracks in the $i$-th bin with reconstructed length $x \in \left[ x_i^{min}, x_i^{\rm max}\right]$ in a sample of mass $M$ which has been recording tracks for a time $t_{\rm age}$ is

\begin{equation} \label{eq:Ntracks}
   N_i = M \times t_{\rm age} \int dx'\; W(x'; x_i^{\rm min}, x_i^{\rm max}) \frac{dR}{dx}(x') \;,
\end{equation}
where $dR/dx$ is given by Eq.~\eqref{eq:dRdx} and we have assumed that $dR/dx$ is independent of time.\footnote{For time-dependent signals, replace $t_{\rm age} \to \int_0^{t_{\rm age}} dt$ in Eq.~\eqref{eq:Ntracks}.} The {\it window function}, $W$, accounts for the finite resolution effects of the readout. In our numerical calculations, we will assume that the probability of measuring a track length $x$ from a track with true length $x'$ is Gaussian-distributed with variance $\sigma_x^2$. The corresponding window function is

\begin{equation}
   W(x'; x_i^{\rm min}, x_i^{\rm max}) = \frac{1}{2} \left[ {\rm erf}\left( \frac{x' - x_i^{\rm min}}{\sqrt{2} \sigma_x} \right) - {\rm erf}\left( \frac{x' - x_i^{\rm max}}{\sqrt{2} \sigma_x} \right) \right] \;.
\end{equation}
Note that in order to avoid artificial sensitivity to tracks which are much shorter than the spatial resolution of the readout, we remove all tracks with true track length $x' < \sigma_x/2$ from the spectra ($dR/dx$) before applying Eq.~\eqref{eq:Ntracks}. For all numerical results in this work, we use 100 log-spaced bins between $\sigma_x/2$ and $1000\,$nm.

\begin{figure}
   \includegraphics[width=0.49\linewidth]{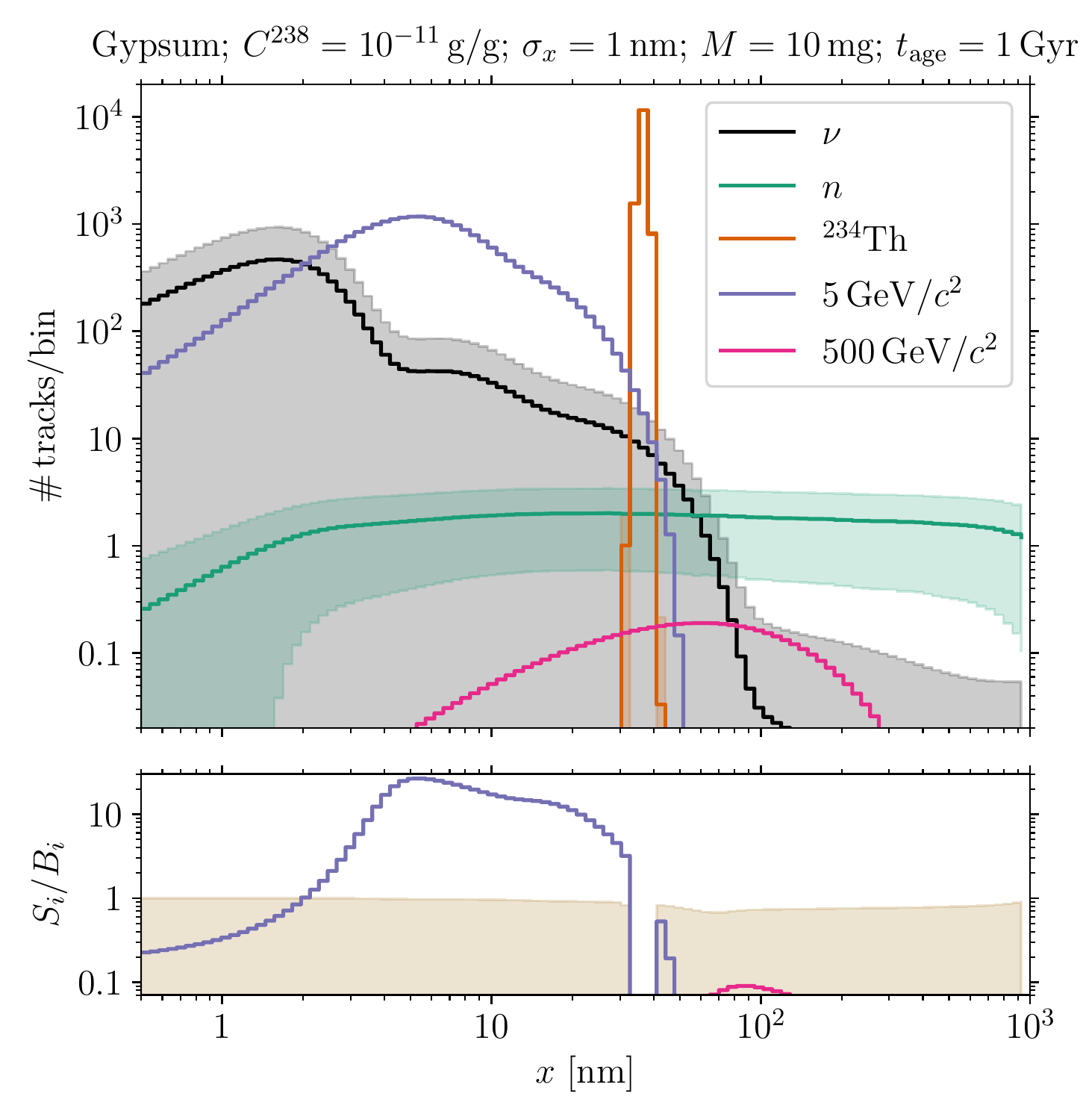}
   \includegraphics[width=0.49\linewidth]{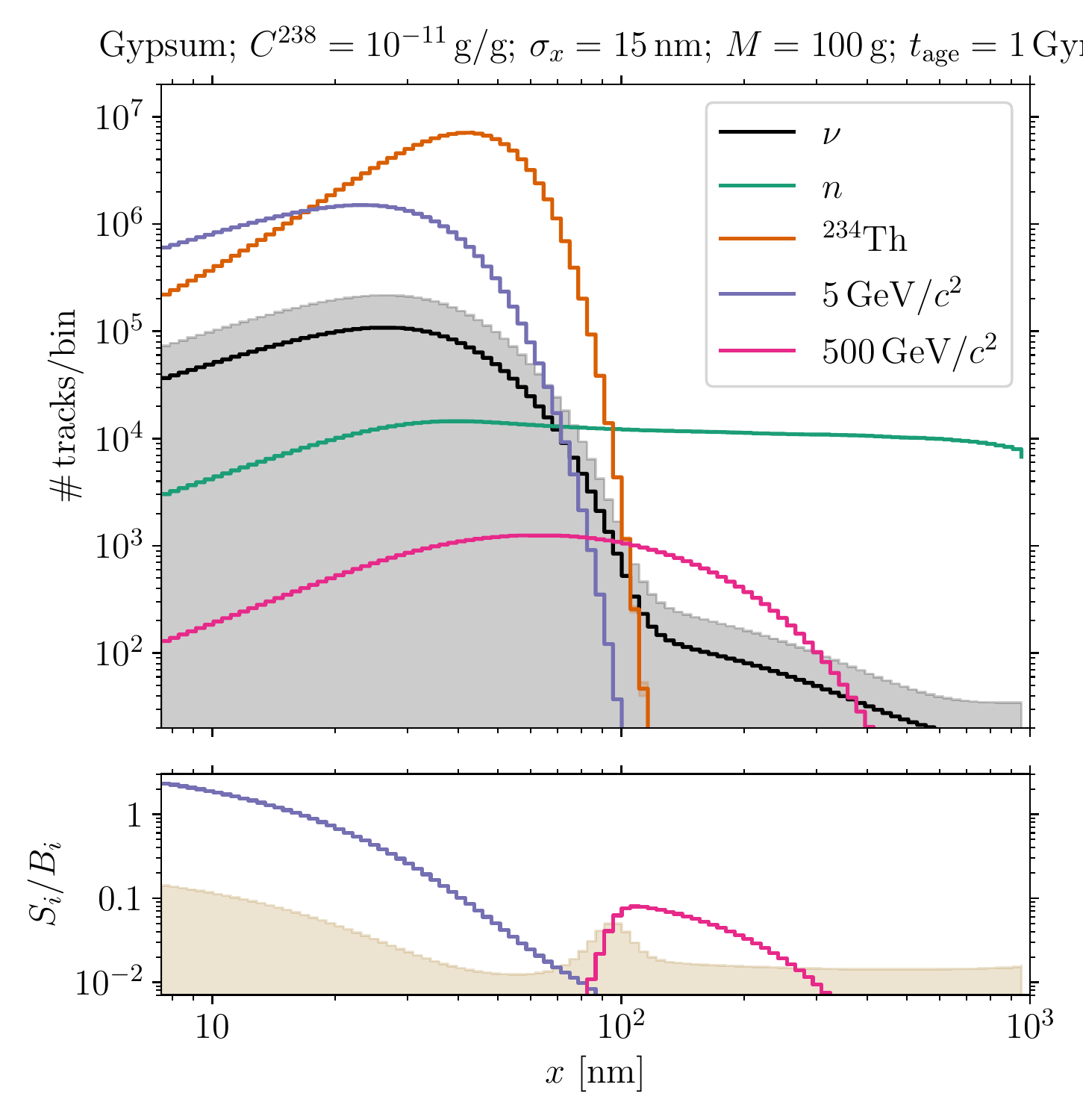}
   \caption{Track length spectra for a 5\,GeV/$c^2$ and a 500\,GeV/$c^2$ WIMP compared to the background spectra induced by neutrinos ($\nu$), radiogenic neutrons ($n$), and $\Ur \to \Th + \alpha$ recoils (\Th) after taking finite resolution effects into account. The left panels are for our {\it high resolution} scenario ($M = 10\,$mg of sample material read out with track length resolution of $\sigma_x = 1\,$nm) while the right panels are for the {\it high exposure} scenario ($M = 100\,$g, $\sigma_x = 15\,$nm). We assume that the samples have been recording tracks for $t_{\rm age} = 1\,$Gyr. In all panels, we are using 100 logarithmically spaced bins between $\sigma_x/2 \leq x \leq 10^3\,$nm; note that these bins and the scales of the axes differ between the left and the right panels. The upper panels show the number of tracks per bin from the respective sources. For the backgrounds, the shaded bands around the respective spectra show the associated error (systematic and statistical added in quadrature). The bottom panels show the ratio of the number of signal ($S_i$) to the (summed) number of background ($B_i$) events per bin, and the sand-colored shade shows the relative uncertainty of the total number of background events per bin. The {\it signal-to-noise} ratio per bin can be read off from the lower panels by dividing $S_i/B_i$ for the respective signal with the relative uncertainty of the background. For the normalization of the WIMP DM spectra we set the spin-independent WIMP-nucleon cross sections to $\sigma_p^{\rm SI} = 10^{-43}\,{\rm cm}^2$ for the 5\,GeV/$c^2$ case and $\sigma_p^{\rm SI} = 10^{-46}\,{\rm cm}^2$ for the 500\,GeV/$c^2$ case, as in Figure~\ref{fig:RangeSpec}. }
   \label{fig:sab}
\end{figure}

Figure~\ref{fig:sab} shows the same 5\,GeV/$c^2$ and 500\,GeV/$c^2$ WIMP DM spectra as the right panel of Figure~\ref{fig:RangeSpec} together with the various background sources after taking into account finite resolution effects. The left panel is for the high-resolution scenario, while the right panel is for the high-exposure scenario. We will discuss this figure further at the end of Section~\ref{sec:Bkg} after introducing the relevant background sources. Let us here only note that the high-resolution scenario is better suited for relatively light WIMPs with masses $m_\chi \lesssim 10\,$GeV/$c^2$, while the high-exposure scenario is aimed at heavier WIMPs, $m_\chi \gtrsim 10\,$GeV/$c^2$. This is for two reasons: First, lighter WIMPs give rise to softer nuclear recoils, resulting in (on average) shorter signal tracks, making excellent track length resolution crucial to resolving the details of the signal and background spectra in the region where the ratio of signal to background events is largest for such light WIMPs. Second, current bounds from direct detection experiments rule out much smaller cross sections for heavier WIMPs with $m_\chi \gtrsim 10\,$GeV/$c^2$, hence, searches for WIMPs with $m_\chi \gtrsim 10\,$GeV/$c^2$ and cross sections smaller than current upper limits require very large exposures. 

\section{Backgrounds} \label{sec:Bkg}

The background sources for DM searches with paleo-detectors are the same as for conventional direct detection experiments. However, the relative importance of the respective sources is different because, compared to conventional direct detection experiments, paleo-detectors use much smaller target masses ($\lesssim 1\,$kg) and integrate over much longer timescales ($t_{\rm age} \sim 0.1 - 1\,$Gyr) without information about the timing of individual signal or background events. The experimental observables in paleo-detectors are nuclear damage tracks with, for all practical purposes, perfect rejection of electronic recoils. Furthermore, natural defects in minerals are either single-site or span across the entire (mono)-crystalline volume and do not resemble nuclear damage tracks. Hence, the only relevant backgrounds are nuclear recoils. In the remainder of this section we will give a brief overview of the most important background sources; we refer the reader to Refs.~\cite{Drukier:2018pdy,Baum:2019fqm} for more detailed discussions.
\begin{itemize}
   \item {\it Cosmogenics:} In order to mitigate backgrounds from cosmic rays, minerals to be used as paleo-detectors must have been shielded by a large overburden for as long as they have been recording nuclear damage tracks (just as conventional direct detection experiments are operated in deep underground laboratories). Paleo-detectors require only (at most) a few kg of target material rather than actively operating a tonne-scale experiment; such modest amounts of materials can be sourced from even greater depths than those of existing underground laboratories. One promising source of samples are (existing) boreholes drilled for geological R\&D or oil exploration. For example, for an overburden of 5\,km rock, the cosmogenic-muon-induced neutron flux is $\mathcal{O}(10^2)\,{\rm cm}^{-2}\,{\rm Gyr}^{-1}$~\cite{Mei:2005gm}, leading to negligible backgrounds for the purposes of a paleo-detector. We note that at depths $\gtrsim 6\,$km rock, backgrounds from atmospheric neutrinos producing neutrons in the vicinity of the target become comparable to the backgrounds from cosmogenic muons~\cite{Aharmim:2009zm}. We also note that paleo-detector samples can be stored near the surface after extraction from deep in the Earth before readout. For example, the cosmogenic-muon-induced neutron flux in a $\sim$50\,m deep storage facility is $\lesssim 0.2\,{\rm cm}^{-2}\,{\rm yr}^{-1}$.

   \item {\it Astrophysical Neutrinos:} Neutrinos scattering off nuclei in a paleo-detector sample give rise to nuclear damage tracks. We include neutrinos from the Sun, supernovae, and the interactions of cosmic rays with the Earth's atmosphere in our background modeling. We take the solar and atmospheric neutrino fluxes from Ref.~\cite{OHare:2020lva}. Because of the long integration times, paleo-detectors are sensitive not only to neutrinos from supernovae in far-away galaxies throughout our Universe (the Diffuse Supernova Neutrino Background, DSNB), but also to neutrinos from local supernovae - the supernova rate in the Milky Way is estimated to be 2--3 per 100\,years~\cite{Cappellaro:2003eg,Diehl:2006cf,Strumia:2006db,Li:2011,Botticella:2011nd,Adams:2013ana}. We compute the DSNB spectrum and the contribution from Galactic supernovae as in Ref.~\cite{Baum:2019fqm}, see also Refs.~\cite{Beacom:2010kk,Adams:2013ana,Billard:2013qya,Madau:2014bja,Strolger:2015kra}. The neutrino-induced background is dominated by solar neutrinos at track lengths below $\sim$100\,nm, with supernova neutrinos giving the largest contributions for tracks of a few 100\,nm, before atmospheric neutrinos dominate the neutrino-induced backgrounds at even longer track lengths. Note that while neutrino-induced nuclear recoils are a background for DM searches, they can also be an interesting signal for paleo-detectors, see Refs.~\cite{Baum:2019fqm,Jordan:2020gxx,Arellano:2021jul}. 

   \item {\it Radiogenics:} Any natural mineral used as a paleo-detector will contain trace amounts of radioactive materials. In order to mitigate radiogenic backgrounds, it is crucial to use minerals with as low a concentration of radioactive elements as possible. The most important radioactive isotope for paleo-detectors is \Ur. Typical minerals formed in the Earth's crust have \Ur concentrations of order $C^{238} \sim 10^{-6}\,$g/g, leading to prohibitively large backgrounds for DM searches. However, minerals which constitute so-called Ultra Basic Rocks (UBRs), formed from the material of the Earth's mantle, and Marine Evaporites (MEs), formed from evaporated sea water, have typical \Ur concentrations orders of magnitude lower~\cite{Thomson:1954,Condie:1957,Adams:1959,Seitz:1974,Dean:1978,Yui:1998,Sanford:2013}, (see also Ref.~\cite{Drukier:2018pdy} and especially the discussion in the appendix of Ref.~\cite{Baum:2019fqm}). As in previous works on paleo-detectors~\cite{Baum:2018tfw,Drukier:2018pdy,Edwards:2018hcf,Baum:2019fqm,Jordan:2020gxx} we will assume benchmark \Ur concentrations of $C^{238} = 10^{-10}\,$g/g for UBRs and $C^{238} = 10^{-11}\,$g/g for MEs. 

   Since paleo-detectors record only nuclear recoils, the most relevant radiogenic background arises from $\alpha$-decays and spontaneous fission. Note that it is not the $\alpha$-particles themselves or the fission fragments which lead to backgrounds for DM searches: $\alpha$-particles (i.e. He nuclei) do not give rise to lasting damage tracks in most materials and, even if they would, the range of the $\alpha$'s is of order $10\,\mu$m in natural minerals, orders of magnitude larger than the tracks from DM-induced nuclear recoils. Similarly, the fission fragments from spontaneous fission of heavy elements have typical energies of tens of MeV, producing much longer tracks than DM-induced recoils. However, in an $\alpha$-decay, the decaying nucleus recoils with typical energies of some tens of keV against the $\alpha$-particle. Due to the long integration time of paleo-detectors, almost all of the \Ur nuclei in the mineral which undergo the first ($\Ur \to \Th + \alpha$) decay in the \Ur decay chain will have continued on to decay to the stable $^{206}$Pb. There are eight $\alpha$-decays in the \Ur decay chain, leading to a characteristic pattern of connected nuclear recoil tracks. We expect such patterns of connected tracks to be easily distinguishable from the isolated recoil tracks DM-nucleus interactions would cause and, hence, do not consider them a relevant background for DM searches. However, the second $\alpha$-decay in the \Ur decay chain ($^{234}{\rm U} \to {^{230}{\rm Th}} + \alpha$) has a relatively long half-life, $T_{1/2}(^{234}{\rm U}) = 0.25\,$Myr. This leads to a population of isolated tracks from \Ur atoms which have undergone the initial ($\Ur \to \Th + \alpha$) decay, but not the second $\alpha$-decay ($^{234}{\rm U} \to {^{230}{\rm Th}} + \alpha$) in the uranium series~\cite{Collar:1995aw,SnowdenIfft:1996zz}. The \Th nucleus receives $72\,$keV of recoil energy in the ($\Ur \to \Th + \alpha$) decay, leading to a population of monochromatic isolated tracks with approximate number density $n(\Th) \simeq 10^6\,{\rm g}^{-1} \times \left( C^{238} / 10^{-11}\,{\rm g/g} \right)$~\cite{Drukier:2018pdy}.

   The second important source of radiogenic backgrounds are neutrons: spontaneous fission of heavy isotopes as well as ($\alpha,n$)-reactions between the $\alpha$-particles produced in $\alpha$-decays and nuclei comprising the target material give rise to neutrons with MeV energies. These neutrons in turn scatter off the nuclei in the paleo-detector sample, giving rise to a broad spectrum of nuclear recoil tracks. In order to model this background, we use \texttt{SOURCES-4A}~\cite{sources4a:1999} to compute the neutron spectra from spontaneous fission and $(\alpha,n)$-reactions of $\alpha$'s from all $\alpha$-decaying nuclei in the \Ur decay chain; note that the contributions from spontaneous fission and ($\alpha,n$)-reactions to the neutron flux are typically of the same order of magnitude, the precise balance depends on the chemical composition of the target. From the neutron spectra, we calculate the induced nuclear recoil spectra using \texttt{TENDL-2017}~\cite{Koning:2012zqy,Rochman:2016,Sublet:2015,Fleming:2015} neutron-nucleus cross sections tabulated in the \texttt{JANIS4.0}~\cite{Soppera:2014zsj} database.\footnote{We take only elastic neutron-nucleus scattering into account; this yields a conservative estimate of the background because neutrons typically lose a larger fraction of their energies through inelastic processes than in elastic scattering. Note also that our Monte Carlo simulation of the nuclear recoils induced by radiogenic neutrons has been validated with a calculation of the nuclear recoils induced by the same neutron spectra with \texttt{FLUKA}~\cite{Ferrari:2005zk,Bohlen:2014buj,NUNDIS} in Ref.~\cite{Jordan:2020gxx} for the particular case of a halite target.} The neutron-induced backgrounds are strongly suppressed in target materials containing hydrogen: since neutrons and H nuclei have approximately the same mass, neutrons lose a large fraction of their energy in a single interaction with hydrogen, leading to efficient moderation of the radiogenic neutrons even if hydrogen comprises only a small fraction of the atoms in the target material~\cite{Baum:2018tfw,Drukier:2018pdy}.
\end{itemize}

In Figures~\ref{fig:RangeSpec} and~\ref{fig:sab} we show the track-length spectra produced by these various background sources together with two benchmark DM signals in gypsum [\ccGyp]. At short track lengths, the background budget is dominated by (solar) neutrino induced nuclear recoils. At intermediate ranges between a few tens and a hundred nm (the precise values depend on the particular mineral), the isolated \Th tracks from $\Ur \to \Th + \alpha$ decays contribute, before the background budget becomes dominated by radiogenic-neutron-induced backgrounds at larger track lengths. Comparing to the benchmark WIMP spectra, we see that solar neutrinos are the most relevant background source for WIMPs with masses in the few GeV/$c^2$ range, while radiogenic neutrons are the limiting background in searches for heavier WIMPs with masses larger than $\sim$10\,GeV/$c^2$.

In Figure~\ref{fig:sab} we show the (binned) number of tracks one would observe in the high-resolution (left panel) and high-exposure (right panel) scenarios after accounting for the effects of finite readout resolution. Because of the large exposure of paleo-detectors, a sizable number of background events will be present. A DM search with paleo-detectors would thus proceed quite differently from a traditional direct detection experiment, in which one typically attempts to construct a signal region with very few (or, ideally, zero) background events, and then searches for an $\mathcal{O}(1)$ number of DM-induced events. Given the large number of background events, in a paleo-detector analysis one would model the distribution of background events (for example, in track-length space) and search for deviations of the observed data from the background-only prediction. Thus, the {\it uncertainty} in the background prediction is much more important than the absolute number of background events. 

For illustration, we include error bands around the respective background components in Figure~\ref{fig:sab}. These bands include both the (statistical) Poisson noise and a {\it systematic} error in the background prediction. For the latter, we make well-motivated assumptions in order to illustrate how the sensitivity of paleo-detectors to signals from DM of various masses would depend on different background components in a cut-and-count analysis, see previous studies of paleo-detectors~\cite{Baum:2018tfw,Drukier:2018pdy,Edwards:2018hcf,Baum:2019fqm,Jordan:2020gxx}. However, we note that in the spectral analysis discussed in Section~\ref{sec:Sens}, the sensitivity of paleo-detectors to a WIMP DM signal is virtually independent of any prior knowledge of the background normalizations. Let us briefly describe our choices for the systematic errors in Figure~\ref{fig:sab}. Neutrino-induced backgrounds are controlled by sources which are not associated with the environment of the paleo-detector target sample: the fluxes of solar, supernova, or atmospheric neutrinos. These fluxes may vary by an $\mathcal{O}(1)$ factor over the $\sim\,$Gyr integration time of paleo-detectors~\cite{Baum:2019fqm,Jordan:2020gxx,Arellano:2021jul}, hence, we assign a (relative) systematic uncertainty of $\pm 100\,\%$ to the neutrino-induced backgrounds.\footnote{In Figure~\ref{fig:sab} we plotted all neutrino-induced backgrounds together to avoid excessive clutter. In our sensitivity projections we treat the solar, DSNB, Galactic supernova, and atmospheric neutrino induced backgrounds as four independent background contributions.} The radiogenic backgrounds, on the other hand, are controlled by the \Ur concentrations which do not change over time. Furthermore, the shape of the track length spectra induced by radiogenic backgrounds can be calibrated by studying the track length distributions in mineral samples with large concentrations of \Ur. Finally, the \Ur concentration in a given sample can be measured using, for example, mass spectrometry or gamma ray spectroscopy~\cite{Povinec:2018,Povinec:2018wgd}. Hence, we expect the uncertainty on the radiogenic background predictions to be much smaller and assign them a (relative) systematic error of $\pm 1\,\%$.

From the bottom panels in Figure~\ref{fig:sab} we can directly read off the {\it signal-to-noise} ratio (per bin), i.e. the ratio of the number of signal events to the uncertainty of the background prediction: in these panels, the lines show the ratio of the number of signal to the (total) number of background events ($S/B$), while the sand-colored shaded region shows the relative uncertainty of the background prediction. For a given signal, the ratio of $S/B$ to the (relative) background uncertainty directly corresponds to the significance one would obtain in a simple cut-and-count analysis. As discussed in Refs.~\cite{Baum:2018tfw,Drukier:2018pdy}, any signal for which $S/B$ is larger than the shaded region indicating the background uncertainty in the bottom panels of Figure~\ref{fig:sab} in at least one bin could be probed with a simple cut-and-count analysis.

Given the large number of events expected in a paleo-detector, one can go beyond the simple cut-and-count analysis and perform a {\it spectral analysis} in track-length space: Armed with predictions for the spectral shape of the various background components and of a possible signal, one can fit all of these spectral templates to the data and ask if the data is better described by either the backgrounds-only or the backgrounds+signal hypothesis. Heuristically, such a search uses {\it control regions}, where some background dominates, to measure that background component, and uses this information to reduce the error of the background prediction in the {\it signal region}, where the signal-to-background ratio is largest. For example, short track lengths can serve as a control region for the solar-neutrino-induced backgrounds (as well as the \Th tracks in the high-exposure scenario), while track lengths longer than $\sim$500\,nm can provide an excellent control region for the neutron-induced backgrounds. We have first used such a spectral analysis in Ref.~\cite{Edwards:2018hcf} and will discuss an alternative approach for the spectral analysis in the next section.

\section{Sensitivity Forecasts} \label{sec:Sens}

\begin{figure}
   \includegraphics[width=0.49\linewidth]{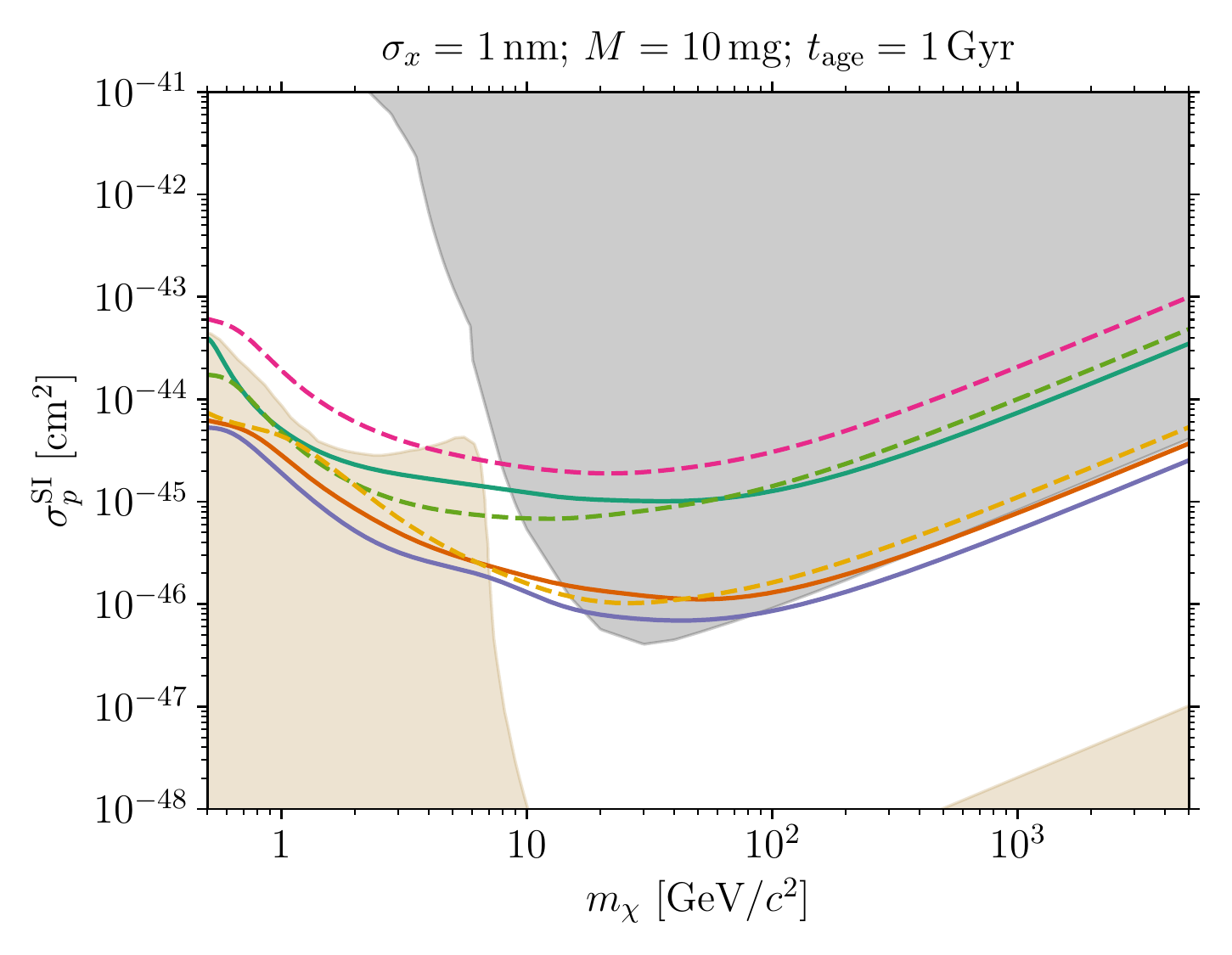}
   \includegraphics[width=0.49\linewidth]{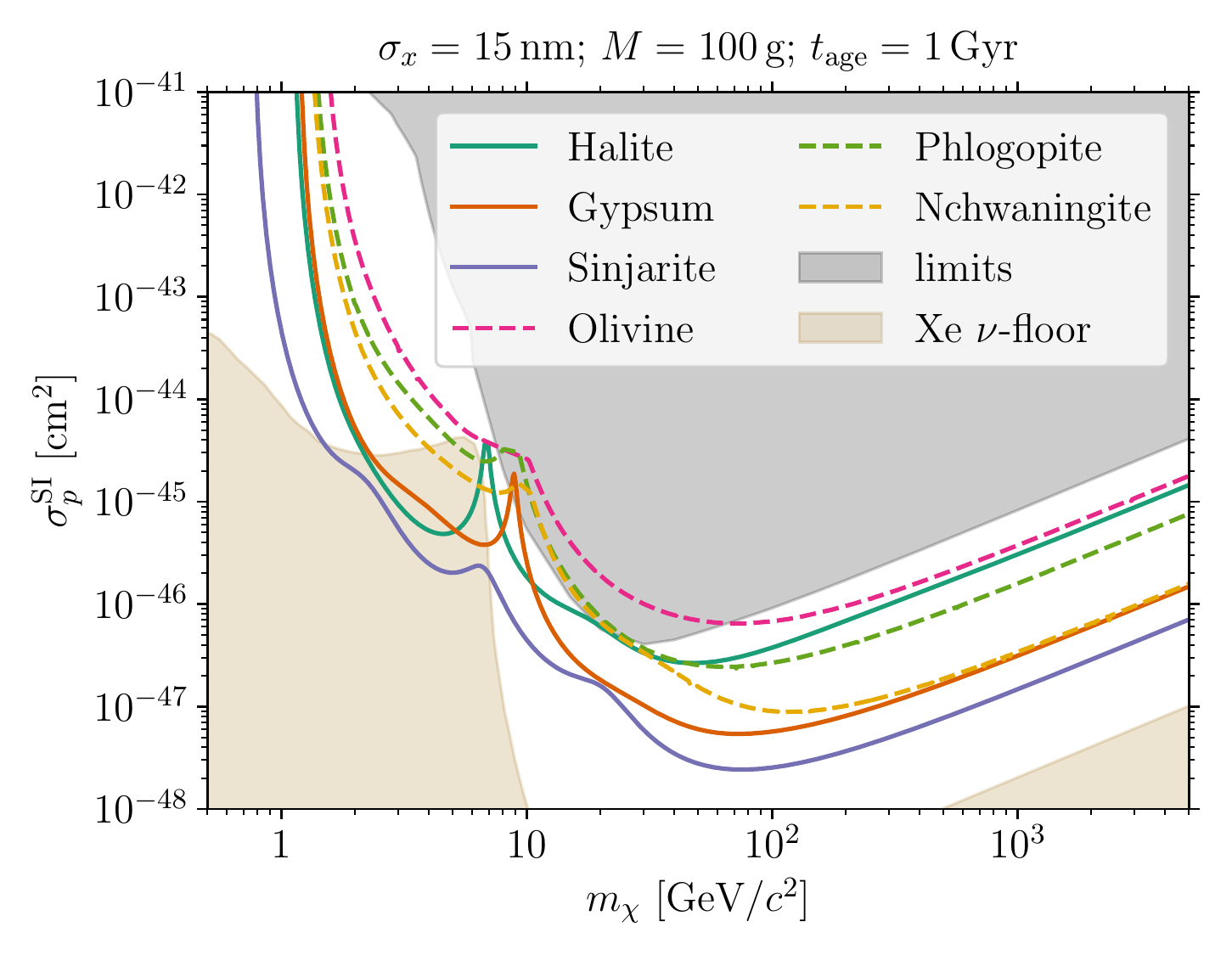}
   \caption{Projected 90\,\% confidence level upper limits in the WIMP mass ($m_\chi$) -- spin-independent WIMP-nucleus scattering cross section ($\sigma_p^{\rm SI}$) plane in the high-resolution (sample mass $M=10\,$mg, track length resolution $\sigma_x = 1\,$nm; left panel) and high-exposure ($M = 100\,$g, $\sigma_x = 15\,$nm; right panel) readout scenarios. The different lines are for different target materials as indicated in the legend, see Table~\ref{tab:Minerals}. The gray-shaded region of parameter space is disfavored by current upper limits from direct detection experiments~\cite{Angloher:2015ewa,Petricca:2017zdp,Agnes:2018ves,Aprile:2018dbl,Aprile:2019xxb}, while the sand-colored region indicates the {\it neutrino floor} for a Xe-based experiment~\cite{Ruppin:2014bra}. Colors and linestyles are the same in both panels.}
   \label{fig:reach_default}
\end{figure}

\begin{figure}
   \includegraphics[width=0.49\linewidth]{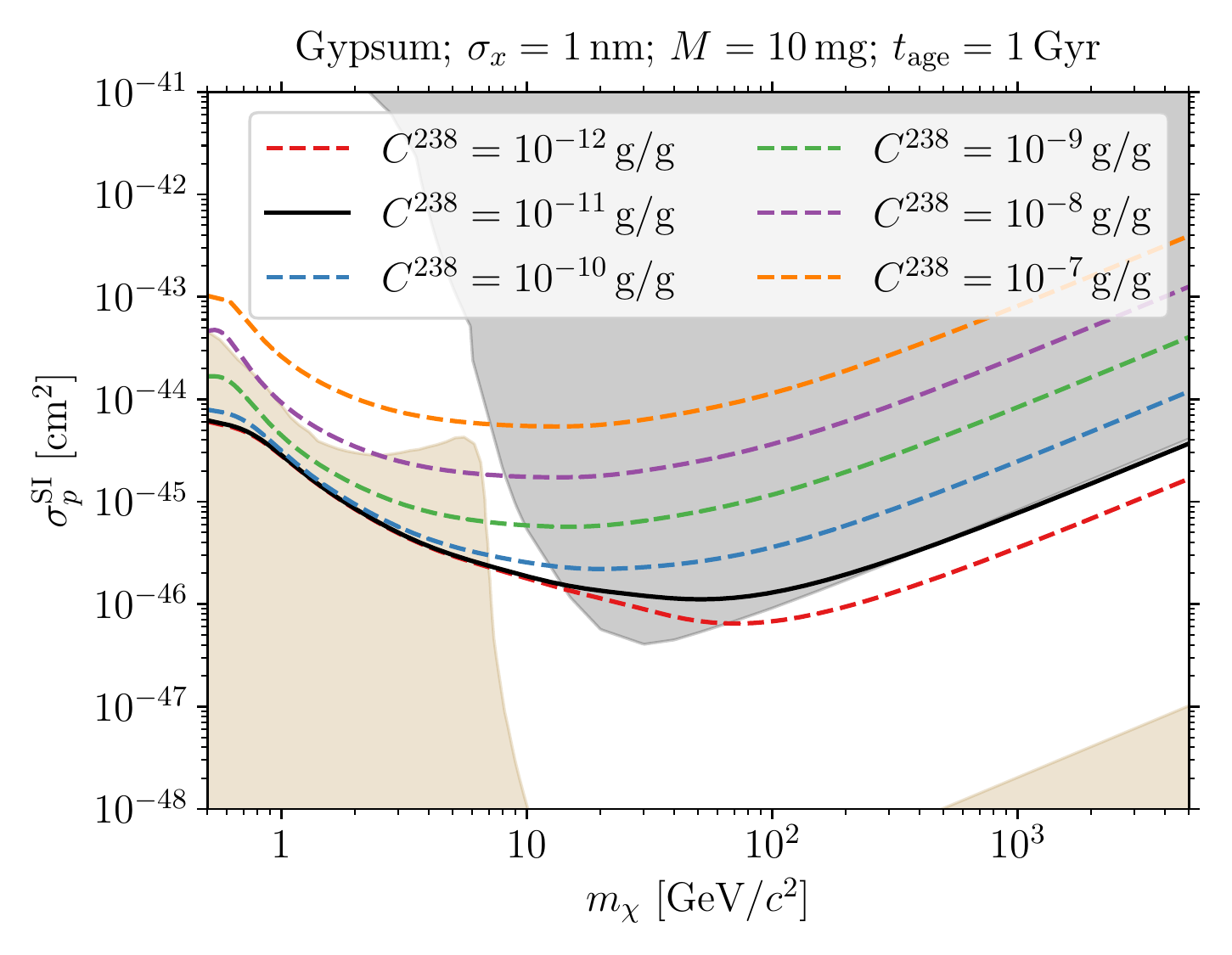}
   \includegraphics[width=0.49\linewidth]{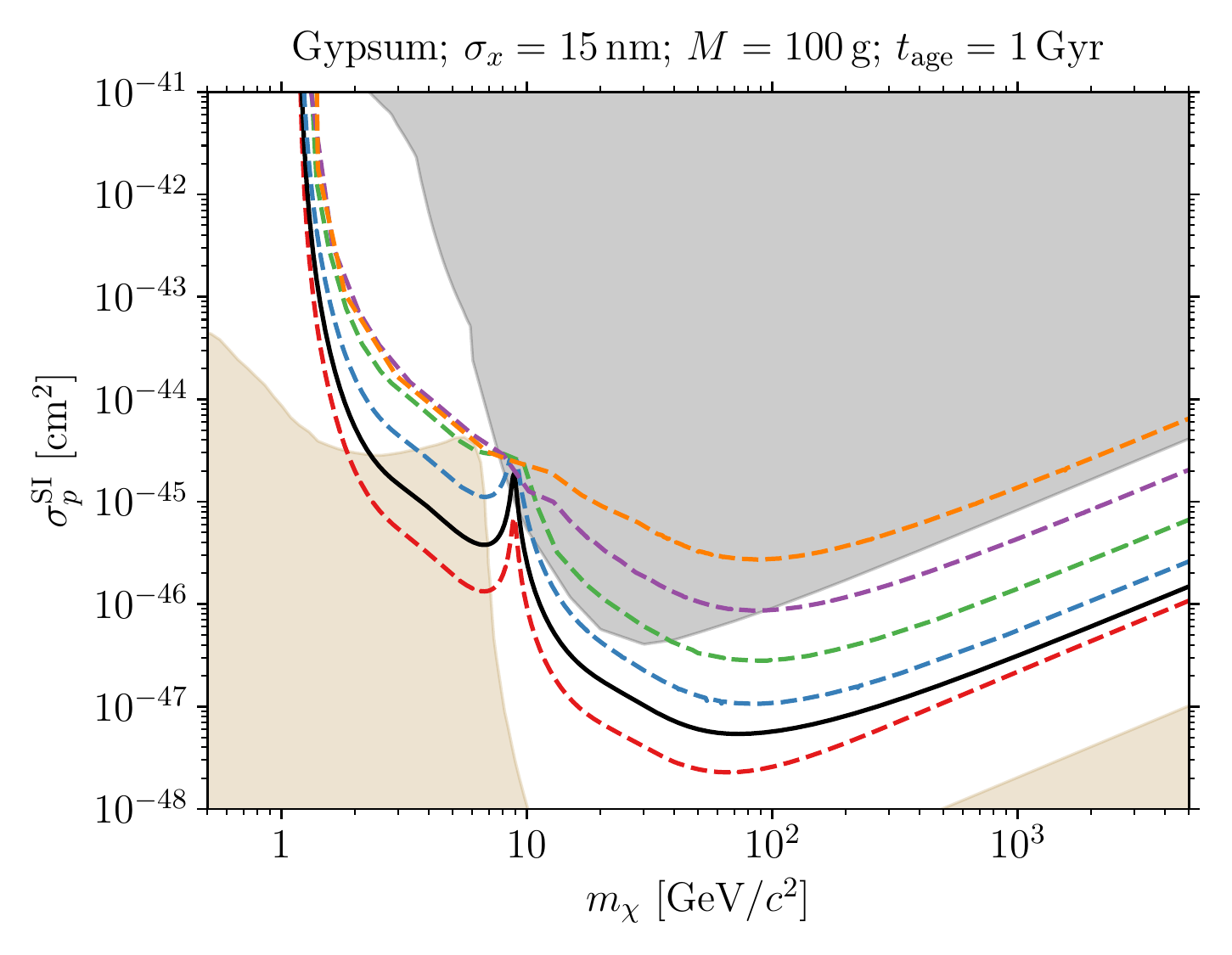}
   \caption{Projected 90\,\% confidence level exclusion limits in a gypsum [\ccGyp] paleo-detector in the high-resolution (left) and high-exposure (right) readout scenarios for different assumptions on the \Ur concentration $C^{238}$ controlling the radiogenic backgrounds. The solid black line shows the sensitivity with our fiducial value for marine evaporites, $C^{238} = 10^{-11}\,$g/g, while the differently colored dashed lines show the projected upper limit for both larger and smaller $C^{238}$. The gray and sand-colored shaded areas indicate current upper limits and the conventional neutrino floor in a Xe-based direct detection experiment, respectively, see Figure~\ref{fig:reach_default}. Colors and linestyles are the same in both panels.}
   \label{fig:reach_C238}
\end{figure}

In Section~\ref{sec:DMsig} we described how to compute the DM signal in paleo-detectors. In particular, we showed how to obtain the track length spectrum one would observe in a paleo-detector from the differential recoil rate per recoil energy and how to account for finite track-length resolution effects. In Section~\ref{sec:Bkg} we discussed the most relevant background sources and briefly sketched how one would perform a DM search with a paleo-detector. Due to the potentially large exposure, a sizable number of background (and, possibly, signal) events will be present in any paleo-detector sample. In order to search for DM, one would perform a spectral analysis of the observed recoil spectrum in track-length space and ask if that spectrum is better fit by a background-only model or by a background+signal hypothesis. In Refs.~\cite{Baum:2018tfw,Drukier:2018pdy} we have used a much simpler cut-and-count analysis to project the sensitivity of paleo-detectors to DM, while in Ref.~\cite{Edwards:2018hcf} we have estimated the sensitivity using a spectral analysis based on the Fisher-matrix formalism using the \texttt{swordfish} package~\cite{Edwards:2017mnf,Edwards:2017kqw,swordfish}. In the remainder of this section, we describe a profile likelihood ratio based approach to calculate the projected sensitivity of paleo-detectors to a DM signal with a spectral analysis. As we will see, one advantage of using this statistical approach is that we can easily incorporate the effect of background mismodeling into our sensitivity forecast. We show results for the sensitivity forecasts in Figures~\ref{fig:reach_default}\,--\,\ref{fig:reach_C238}. 

The predicted track length spectrum in a paleo-detector is 

\begin{equation} \label{eq:TotalSpec} \begin{split}
   {\bm N}({\bm \theta}; \sigma_p^{\rm SI}, m_\chi) &= {\bm N}_\nu^{\rm sol}(\Phi_\nu^{\rm sol}) + {\bm N}_\nu^{\rm DSNB}(\Phi_\nu^{\rm DSNB}) + {\bm N}_\nu^{\rm GSNB}(\Phi_\nu^{\rm GSNB}) + {\bm N}_\nu^{\rm atm}(\Phi_\nu^{\rm atm}) \\
   &\quad + {\bm N}_{\rm rad}^{\Th}(C^{238}) + {\bm N}_{\rm rad}^n(C^{238}) \\
   &\quad + {\bm N}^{\rm DM}(\sigma_p^{\rm SI}; m_\chi) \;,
\end{split} \end{equation}
where the ${\bm N}$ are the track length spectra described in Eq.~\eqref{eq:Ntracks}. Each ${\bm N}$ is computed from the appropriate recoil energy distribution, given by Eq.~\eqref{eq:recoilSpec} for the DM signal and the corresponding expressions for the different background components discussed in Section~\ref{sec:Bkg}. In Eq.~\eqref{eq:TotalSpec}, the terms in the first line describe the contributions induced by solar neutrinos (${\bm N}_\nu^{\rm sol}$), the DSNB (${\bm N}_\nu^{\rm DSNB}$), the Galactic Supernova Neutrino Background (${\bm N}_\nu^{\rm GSNB}$), and atmospheric neutrinos (${\bm N}_\nu^{\rm atm}$). The terms in the second line describe the radiogenic backgrounds, separated into: the isolated \Th decays from \Ur nuclei, which have undergone only the inital $\alpha$-decay during the time the sample has been recording nuclear damage tracks (${\bm N}_{\rm rad}^{\Th}$), and the background induced by radiogenic neutrons (${\bm N}_{\rm rad}^n$). Finally, the term in the third line is the DM signal. For each $\bm N$, we have indicated in parantheses the parameter controlling its normalization: the flux ($\Phi_\nu^i$) for each neutrino contribution, the \Ur concentration in the sample ($C^{238}$) for the radiogenic backgrounds, and the WIMP-nucleon cross section ($\sigma_p^{\rm SI}$) for the DM signal. Note that all contributions are also proportional to the mass of the sample, $M$, and all contributions except for ${\bm N}_{\rm rad}^{\Th}$ are proportional to the length of time the sample has been recording damage tracks, $t_{\rm age}$; we have suppressed these dependencies in Eq.~\eqref{eq:TotalSpec} for compactness. Finally, the DM signal also depends on the mass of the WIMP, $m_\chi$. 

To easily compare to existing upper limits and the projected sensitivities of conventional direct detection experiments, we calculate the projected upper (exclusion) limit on $\sigma_p^{\rm SI}$ at 90\,\% confidence level as a function of $m_\chi$ for a paleo-detector. The set of nuisance parameters in Eq.~\eqref{eq:TotalSpec} is then

\begin{equation} \label{eq:Nuisance}
   {\bm \theta} = \left\{ M, t_{\rm age}, \Phi_\nu^{\rm sol}, \Phi_\nu^{\rm DSNB}, \Phi_\nu^{\rm GSNB}, \Phi_\nu^{\rm atm}, C^{238} \right\} \;.
\end{equation} 
Since paleo-detectors are ultimately a counting experiment, we expect the data to be Poisson-distributed. The appropriate log-likelihood to observe the data ${\bm D}$ given the parameters $\left\{ {\bm \theta}; \sigma_p^{\rm SI}, m_\chi \right\}$ is thus\footnote{Here and in the following, we drop constant factors in the expression of the likelihood which cancel in the likelihood ratio we are ultimately interested in.}

\begin{equation} \label{eq:LLpoi}
   \ln {\mathcal L}_{\rm Poisson} \left( {\bm D} \middle| {\bm \theta}; \sigma_p^{\rm SI}, m_\chi \right) = \sum_i \left[ D_i \ln N_i({\bm \theta}; \sigma_p^{\rm SI}, m_\chi) - N_i({\bm \theta}; \sigma_p^{\rm SI}, m_\chi) \right] \;.
\end{equation}
The nuisance parameters will be constrained by other measurements. For example, the neutrino fluxes will be constrained by existing neutrino experiments and theoretical modeling. The mass, age, and uranium concentration of the sample will be constrained by direct measurements of the sample mineral. In a frequentist analysis, these constraints can be taken into account by jointly analyzing these ancillary measurements with the track length spectrum data. To mimick such a joint analysis, we include a set of external (Gaussian) constraints on the nuisance parameters,

\begin{equation} \label{eq:LLext}
   \ln {\mathcal{L}}_{\rm ext.~const.}({\bm \theta}) = \sum_j \left[ - \frac{1}{2} \left( \frac{\theta_j - \overline{\theta}_j}{c_j \overline{\theta}_j} \right)^2 \right] \;.
\end{equation}
Here, $\overline{\theta}_j$ is the central value for the $j$-th nuisance parameter supplied by the ancillary measurement(s), and $c_j$ is the {\it relative} uncertainty of the measurement(s). The likelihood is then 

\begin{equation} 
   \ln {\mathcal L} \left( {\bm D} \middle| {\bm \theta}; \sigma_p^{\rm SI}, m_\chi \right) = \ln {\mathcal L}_{\rm Poisson}  \left( {\bm D} \middle| {\bm \theta}; \sigma_p^{\rm SI}, m_\chi \right) + \ln {\mathcal{L}}_{\rm ext.~const.}({\bm \theta}) \;.
\end{equation}

To calculate the projected sensitivity of paleo-detectors we consider the maximum log-likelihood ratio~\cite{Cowan:2010js,Billard:2012,Conrad:2014nna}

\begin{equation} \label{eq:TS}
   q(\sigma_p^{\rm SI}; m_\chi) = - 2 \ln \left[ \frac{\mathcal{L} \left( {\bm D} \middle| \hat{\hat{\bm \theta}}; \sigma_p^{\rm SI}, m_\chi \right)}{\mathcal{L} \left( {\bm D} \middle| \hat{\bm \theta}; \hat{\sigma}_p^{\rm SI}, m_\chi \right)} \right] \;.
\end{equation}
In the numerator, $\hat{\hat{{\bm \theta}}}$ is the set of nuisance parameters that maximizes $\mathcal{L}$ for given values of $\sigma_p^{\rm SI}$ (and $m_\chi$), while in the denominator, $\mathcal{L}$ is maximized over both ${\bm \theta}$ and $\sigma_p^{\rm SI}$ (for a given value of $m_\chi$), with best-fit values $\hat{\bm \theta}$ and $\hat{\sigma}_p^{\rm SI}$. In order to calculate the projected exclusion limit, we use the {\it Asimov} data set~\cite{Cowan:2010js} under the assumption of no signal,

\begin{equation}
   {\bm D} = {\bm N}(\overline{\bm \theta}; \sigma_p^{\rm SI}=0) \;,
\end{equation}
where $\overline{\bm \theta}$ is a set of fiducial choices for the nuisance parameters. The 90\,\% confidence level exclusion limit for a given value of $m_\chi$ is obtained by finding the smallest value of $\sigma_p^{\rm SI}$ for which $q(\sigma_p^{\rm SI}; m_\chi) \geq q_{\rm crit} = 2.71$.\footnote{By Wilks' theorem~\cite{Wilks:1938dza}, the distribution of $q$ is (asymptotically) $\chi^2$ distributed, and for a one-dimensional $\chi^2$-distribution, $\chi^2 = 2.71$ for a $p$-value of 0.1.} 

We describe the the fiducial values for the nuisance parameters listed in Eq.~\eqref{eq:Nuisance} and the relative uncertainty of the external constraints on those parameters, $c_j$, defined by Eq.~\eqref{eq:LLext}. For the high-resolution (high-exposure) scenario, we choose a fiducial value of $\overline{M} = 10\,$mg ($\overline{M} = 100\,$g) for the sample mass and, for both scenarios, $\overline{t}_{\rm age} = 1\,$Gyr for the time the sample has been recording damage tracks. We assume that these parameters can be constrained by ancillary measurements to $c_M = 0.01\,\%$ and $c_{t_{\rm age}} = 5\,\%$; while measuring the mass of a target sample to high precision is trivial, its age can be inferred with few-percent precision using standard geological dating techniques~\cite{GTS2012,Gallagher:1998,vandenHaute:1998}. For the normalization of the neutrino fluxes, $\overline{\Phi}_\nu^i$, and the \Ur concentrations in our target minerals, $\overline{C}^{238}$, we take fiducial values described in Section~\ref{sec:Bkg} (the latter are also summarized in Table~\ref{tab:Minerals}). As a benchmark, we assume that the associated nuisance parameters can be constrained to $c_{\Phi_\nu^i} = 100\,\%$ and $c_{C^{238}} = 1\,\%$; these choices are equivalent to the respective systematic uncertainties of the backgrounds assumed in previous studies~\cite{Baum:2018tfw,Drukier:2018pdy,Edwards:2018hcf,Baum:2019fqm,Jordan:2020gxx}. However, as we will discuss below, the projected sensitivity of paleo-detectors to a DM signal is largely independent of external constraints on the nuisance parameters. 

\begin{table}
   \begin{tabular}{ccccccc}
      \hline \hline
      \textbf{Mineral}  & \textbf{Composition} & \textbf{Fiducial $^{\bf 238}$U concentration [g/g]}\\
      \hline
      Halite & \ccHal & $10^{-11}$ \\
      Gypsum & \ccGyp & $10^{-11}$ \\
      Sinjarite & \ccSin & $10^{-11}$ \\
      Olivine & \ccOli & $10^{-10}$ \\
      Phlogopite & \ccPhl & $10^{-10}$ \\
      Nchwaningite & \ccNch & $10^{-10}$ \\
      \hline\hline
   \end{tabular}
   \caption{List of minerals considered in this work with their chemical composition and our fiducial assumption for the \Ur concentration ($C^{238}$) in radiopure samples of these minerals.}
   \label{tab:Minerals}
\end{table}

In Figure~\ref{fig:reach_default} we show the projected 90\,\% confidence level upper limit under these fiducial assumptions in both the high-resolution (sample mass $M = 10\,$mg, track-length resolution $\sigma_x = 1\,$nm; left panel) and the high-exposure ($M = 100\,$g, $\sigma_x = 15\,$nm; right panel) readout scenario. We show results for a selection of minerals, see Table~\ref{tab:Minerals}: three examples of Ultra-Basic Rocks (UBRs), the very common olivine and two less common UBRs which contain hydrogen, phlogopite mica and nchwaningite. We also show the projected sensitivity for three examples of Marine Evaporites (MEs): halite, gypsum, and sinjarite where the latter two contain hydrogen. As discussed above, the high-resolution scenario has better sensitivity for WIMP masses $m_\chi \lesssim 10\,$GeV/$c^2$ where the dominant background stems from solar neutrinos; here, excellent track length resolution is crucial to resolving the differences between the shapes of the signal and background track length spectra. For $m_\chi \gtrsim 10\,$GeV/$c^2$, the high-exposure scenario has better sensitivity since for such WIMP masses radiogenic backgrounds are dominant and maximizing the exposure is crucial in order to achieve the best sensitivity. Comparing the sensitivity for the different target materials we can note that materials containing hydrogen generally have better sensitivity than those without hydrogen. This is due to the effective suppression of the radiogenic neutron backgrounds by hydrogen, see the discussion in Section~\ref{sec:Bkg}. Furthermore, we note that MEs typically have better sensitivity than UBRs; this is because we assume a lower \Ur concentration ($C^{238} = 10^{-11}\,$g/g) for MEs than for UBRs ($C^{238} = 10^{-10}\,$g/g). Out of the minerals considered in Figure~\ref{fig:reach_default}, throughout the shown WIMP mass range and for both the high-resolution and the high-exposure readout scenarios, sinjarite promises the best sensitivity. Primarily, this is due to sinjarite being an ME containing hydrogen, and hence having relatively small \Ur-induced backgrounds. However, the sensitivity is also enhanced by its particular chemical composition: the other example of an ME containing hydrogen shown in Figure~\ref{fig:reach_default}, gypsum, is very oxygen-rich, and hence, both the DM-induced and the background spectra are dominated by the contribution of oxygen. Sinjarite on the other hand has one calcium, two oxygen, and two chlorine atoms; each of these elements can therefore contribute comparably to the recoil spectra. This adds additional structure to the track length spectra which is advantageous in a spectral analysis.

In Figure~\ref{fig:reach_default} we also show existing limits from conventional direct detection experiments~\cite{Angloher:2015ewa,Petricca:2017zdp,Agnes:2018ves,Aprile:2018dbl,Aprile:2019xxb} as well as the neutrino floor for a Xe-based experiment~\cite{Ruppin:2014bra} with the shaded regions. At small WIMP masses, $m_\chi \lesssim 10\,$GeV/$c^2$, paleo-detectors could probe WIMPs with WIMP-nucleon cross sections many orders of magnitude smaller than those currently probed by conventional experiments; for example, at $m_\chi = 1\,$GeV/$c^2$ the best existing upper limit is $\sigma_p^{\rm SI} \lesssim 10^{-38}\,{\rm cm}^2$~\cite{Angloher:2015ewa,Petricca:2017zdp}. Moreover, paleo-detectors offer sensitivity deep into the conventional (solar) neutrino floor in the $m_\chi \lesssim 10\,$GeV/$c^2$ mass range. Due to the large exposures, a paleo-detector would observe $\mathcal{O}(100 - 1000)$ recoil tracks from solar neutrinos in the high-resolution readout scenario, see Figure~\ref{fig:sab}. In a spectral analysis, this large number of neutrino-induced events combined with the excellent track-length (and, in turn, recoil energy) resolution would allow one to characterize the solar neutrino background in the ``signal region'' and hence probe WIMP-nucleon cross sections deep into the conventional neutrino floor. For heavier WIMPs ($m_\chi \gtrsim 10\,$GeV/$c^2$), we see that paleo-detectors could probe WIMP-nucleon cross sections roughly one and a half orders of magnitude below current upper limits, but somewhat larger than the conventional neutrino floor in a Xe-based direct detection experiment. In this WIMP mass region, the sensitivity of paleo-detectors is limited by the radiogenic neutron backgrounds, as we will discuss further below.

In the remainder of this section, we will explore the robustness of the sensitivity projections shown in Figure~\ref{fig:reach_default} to the assumptions we have made. Let us first note that the projected exclusion limits are virtually independent of the ancillary measurements. For fixed normalization of the backgrounds in the Asimov data, removing the Gaussian constraints [defined by Eq.~\eqref{eq:LLext}] on the parameters controlling the normalization of the neutrino-induced backgrounds ($\Phi_\nu^i$) and the radiogenic backgrounds ($C^{238}$) as well as removing the external constraint on the age of the samples ($t_{\rm age}$) leads to a depreciation of the projected upper limits by less than a factor two across the WIMP mass range shown in Figure~\ref{fig:reach_default}. We find the most pronounced depreciation of the upper limits in the $m_\chi \lesssim 1\,$GeV/$c^2$ WIMP range when the constraints on the $\Phi_\nu^i$ are removed. Hence, the data from the track-length spectrum alone suffices to search for DM with paleo-detectors, even in the absence of any ancillary information about the normalization of the backgrounds.

\begin{figure}
   \includegraphics[width=0.49\linewidth]{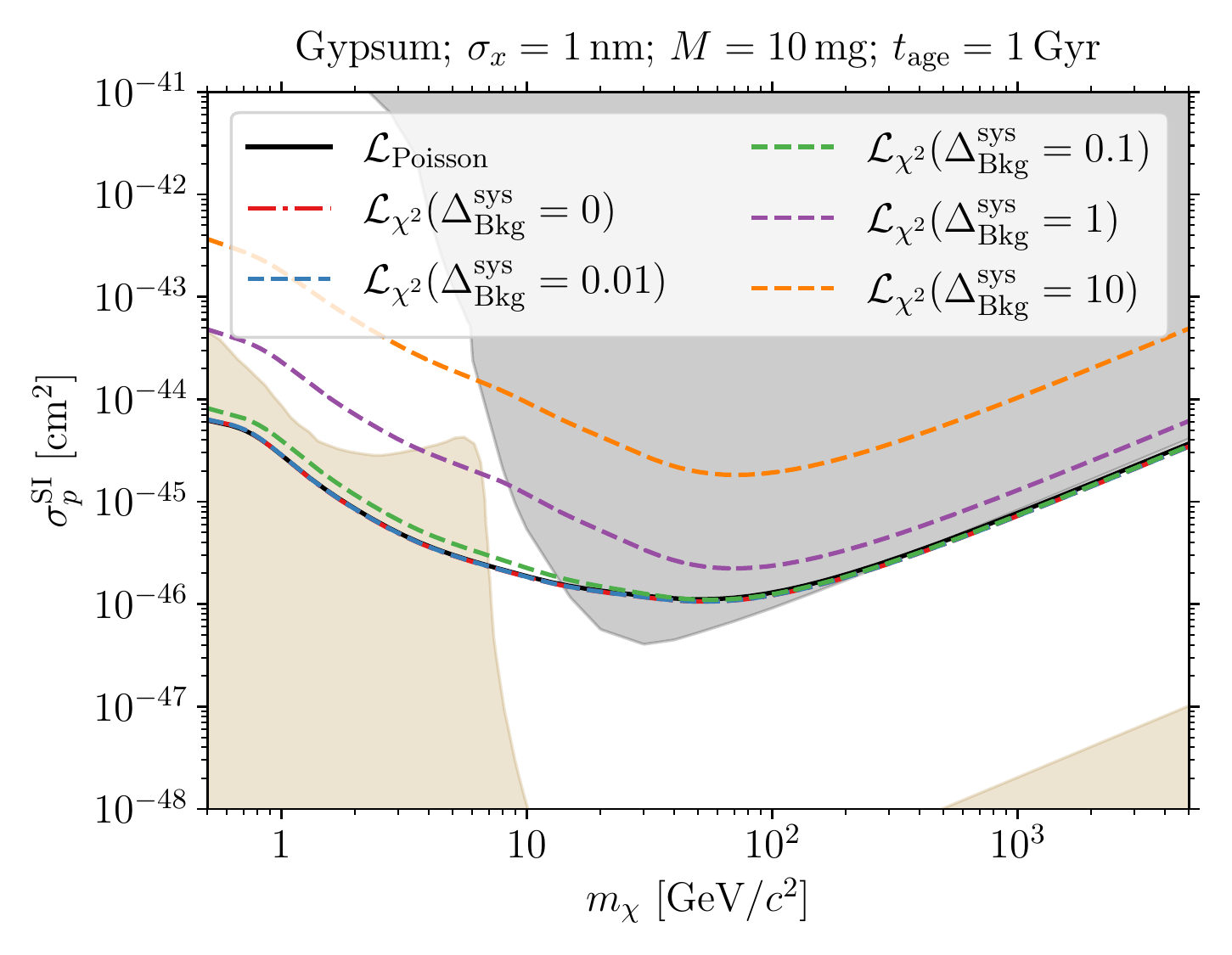}
   \includegraphics[width=0.49\linewidth]{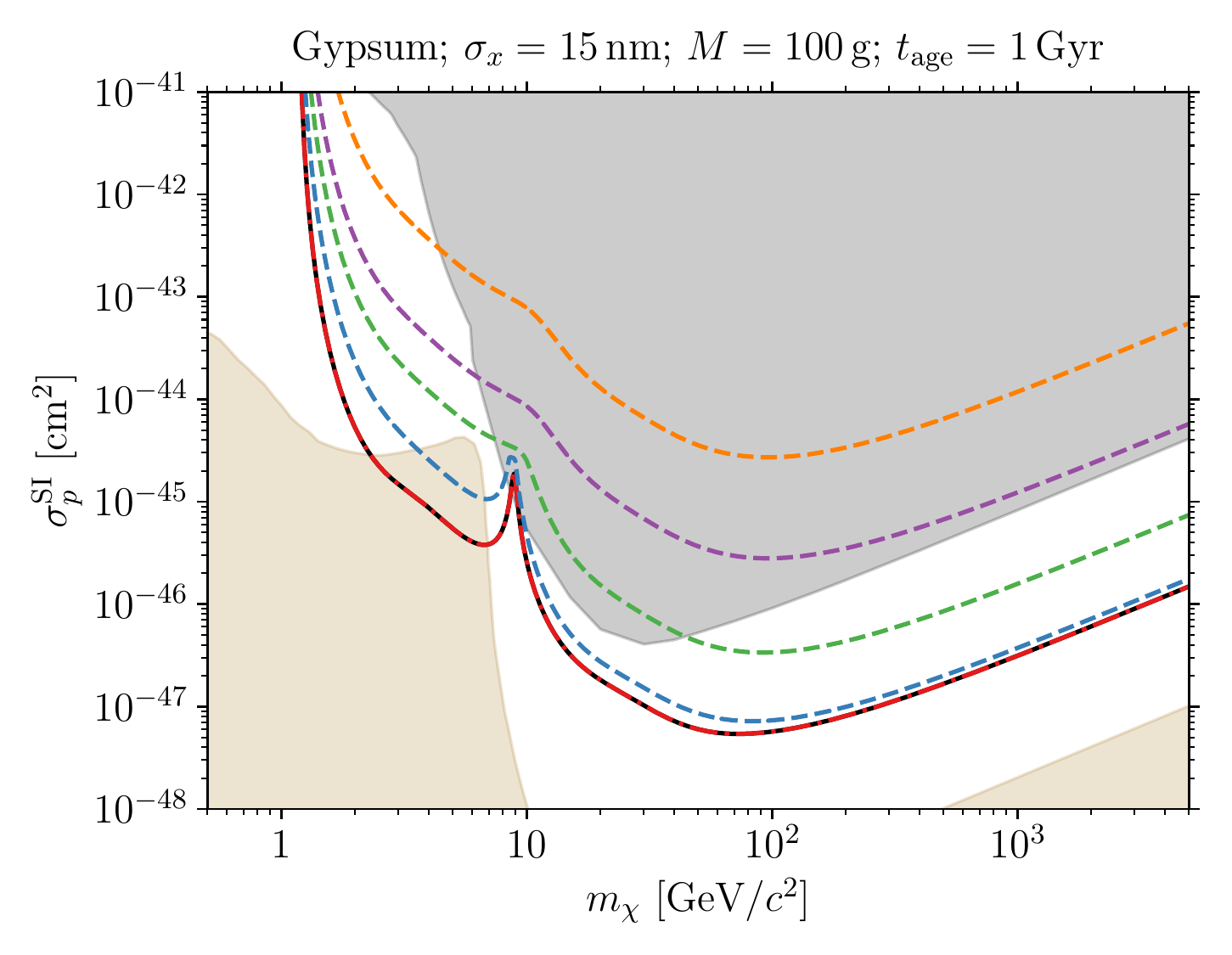}
   \caption{Projected 90\,\% confidence level exclusion limits in a gypsum [\ccGyp] paleo-detector in the high-resolution (left) and high-exposure (right) readout scenarios. The black solid lines show the projected limit using the Poisson log-likelihood, see Eq.~\eqref{eq:LLpoi}, in our maximum likelihood ratio test, Eq.~\eqref{eq:TS}, while the dashed lines show the projected upper limit if we instead use a $\chi^2$ distribution, Eq.~\eqref{eq:LLx2}, for the log-likelihood. The differently colored dashed lines are for different assumptions on the relative systematic error of the background modeling parameterized by $\Delta_{\rm Bkg}^{\rm sys}$, see Eq.~\eqref{eq:x2Var}. The gray and sand-colored shaded areas indicate current upper limits and the conventional neutrino floor in a Xe-based direct detection experiment, respectively, see Figure~\ref{fig:reach_default}. Colors and linestyles are the same in both panels. Note that in the left panel, the $\mathcal{L}_{\rm Poisson}$, the $\mathcal{L}_{\chi^2}(\Delta_{\rm Bkg}^{\rm sys} = 0)$, and the $\mathcal{L}_{\chi^2}(\Delta_{\rm Bkg}^{\rm sys} = 0.01)$ lines are practically laying on top of each other.}
   \label{fig:reach_sys}
\end{figure}

In order to explore how robust the sensitivity forecasts are with respect to mismodeling of the {\it shape} of each background component, we replace the Poisson log-likelihood, Eq.~\eqref{eq:LLpoi}, with a $\chi^2$ distribution,

\begin{equation} \label{eq:LLx2}
   \ln {\mathcal L}_{\chi^2} \left( {\bm D} \middle| {\bm \theta}; \sigma_p^{\rm SI}, m_\chi \right) = \sum_i \left[ - \frac{1}{2} \left( \frac{D_i - N_i({\bm \theta}; \sigma_p^{\rm SI}, m_\chi)}{\sigma_i} \right)^2 \right] \;,
\end{equation}
where $\sigma_i^2$ is the variance in the $i$-th bin. If we include only the Poisson error of the data, $\sigma_i^2 \to D_i$, the Poisson log-likelihood and the $\chi^2$ likelihood become asymptotically equivalent in the large-$N_i$ limit. Given the number of signal events one would observe in a paleo-detector, see Figure~\ref{fig:sab}, we thus expect to find the same sensitivity for both forms of the likelihood. The $\chi^2$ distribution allows us to include additional uncertainties, which we model as

\begin{equation} \label{eq:x2Var}
   \sigma_i^2 = D_i + \left(\Delta_{\rm Bkg}^{\rm sys} D_i\right)^2 \;,
\end{equation}
where $\Delta_{\rm Bkg}^{\rm sys}$ parameterizes the (relative) systematic modeling error. We note that it is straightforward to extend this to $\sigma_i^2 = D_i + \sum_{j} \left( \Delta_i^j D_i^j \right)^2$, allowing for the assignment of a systematic error to the $j$-th contribution to the data in the $i$-th bin. In this work, we will only employ a simple overall systematic error $\Delta$ to explore the robustness of the sensitivity projections. 

In Figure~\ref{fig:reach_sys} we show the projected upper limit for a gypsum paleo-detector under different assumptions for the background-shape mismodeling. First, we note that for $\Delta_{\rm Bkg}^{\rm sys} = 0$, i.e. accounting only for the Poisson error in the variance, $\sigma_i^2 = D_i$, both the Poisson log-likelihood and the $\chi^2$ distribution yield the same projected limit. In the high-resolution scenario (left panel of Figure~\ref{fig:reach_sys}) we find that if we include an error of $\Delta_{\rm Bkg}^{\rm sys} = 0.1$, corresponding to 10\,\% bin-to-bin mismodeling of the shape of the backgrounds, the sensitivity does not depreciate significantly. For $\Delta_{\rm Bkg}^{\rm sys} = 1$, the projected upper limit is up to an order of magnitude worse than in the $\Delta_{\rm Bkg}^{\rm sys} = 0$ case for $m_\chi \lesssim 30\,$GeV/$c^2$, with smaller changes at larger WIMP masses. In the high-exposure scenario (right panel of Figure~\ref{fig:reach_sys}), the sensitivity is not quite as robust against background-shape mismodeling. For small systematic errors, $\Delta_{\rm Bkg}^{\rm sys} = 0.01$, we find that the sensitivity becomes only slightly worse than for $\Delta_{\rm Bkg}^{\rm sys} = 0$, while for $\Delta_{\rm Bkg}^{\rm sys} = 0.1$ the projected upper limit is a factor of $\sim 5$ larger than for $\Delta_{\rm Bkg}^{\rm sys} = 0$. The high-exposure-readout scenario is most relevant for the mass range $m_\chi \gtrsim 10\,$GeV/$c^2$, where the dominant background stems from radiogenic neutrons. The corresponding track length spectrum is straightforward to calibrate in paleo-detectors by measuring the tracks in samples of the same mineral but with larger \Ur concentration. Thus, it seems unlikely that the systematic (bin-to-bin) shape errors would be as large as $\Delta_{\rm Bkg}^{\rm sys} = 0.1$ in an actual paleo-detector experiment. However, we note from Figure~\ref{fig:reach_sys} that even for errors as large as $\Delta_{\rm Bkg}^{\rm sys} = 0.1$, a gypsum paleo-detector could still probe WIMP-nucleon cross sections almost an order of magnitude below current experimental limits for $m_\chi \gtrsim 100\,$GeV/$c^2$. 

In Figure~\ref{fig:reach_C238} we show the effect of assuming different \Ur concentrations on the projected upper limits in a gypsum paleo-detector. Trivially, we find that paleo-detectors could probe smaller WIMP cross sections if the samples are more radiopure. For most of the range of \Ur concentrations, $10^{-12}\,{\rm g/g} \leq C^{238} \leq 10^{-7}\,$g/g shown in Figure~\ref{fig:reach_C238} and most of the shown WIMP mass range, the projected upper limit scales like $\sigma_p^{\rm SI} \propto \sqrt{C^{238}}$. This scaling arises from the uncertainty of the radiogenic backgrounds in the ``signal region'' being given by the associated Poisson error, with the normalization of the radiogenic backgrounds inferred from the relevant ``control regions''. The $\propto \sqrt{C^{238}}$ scaling of the sensitivity is broken if other backgrounds provide significant contributions in the ``signal region.'' For example, the radiogenic background contribution can be subdominant to that of solar neutrinos for $m_\chi \lesssim 10\,$GeV/$c^2$ or, for very small \Ur concentrations, atmospheric neutrinos for $m_\chi \gtrsim 100\,$GeV/$c^2$. Note that in the high-resolution scenario, paleo-detectors could probe WIMP-nucleon cross sections down to the conventional (solar) neutrino floor even for \Ur concentrations as large as $C^{238} = 10^{-8}\,$g/g in the $m_\chi \lesssim 10\,$GeV/$c^2$ mass range. For heavier WIMPs, we find that a gypsum paleo-detector in the high-exposure readout scenario could still probe WIMP-nucleus cross sections below current experimental limits even if the \Ur concentration is orders of magnitude larger than our fiducial assumption of $C^{238} = 10^{-11}\,$g/g. 

Before moving on, let us note that the statistical framework used here to project the sensitivity of paleo-detectors can easily be adapted to answer different statistical questions or to incorporate additional uncertainties. For example, by using the Asimov data as defined above (backgrounds only) but also including a DM signal contribution, one can calculate the projected {\it discovery reach} rather than the exclusion limit. Using the $\chi^2$ likelihood, it is straightforward to include arbitrarily complicated uncertainties; another approach to including (modeling) uncertainties would be to use random realizations of noisy mock data instead of the Asimov data set. It is also straightforward to extend the framework to simultaneously analyze a series of paleo-detectors. Samples of different ages can be used to explore the sensitivity of paleo-detectors to time-dependent signals. For instance, the Solar System could have traversed some denser region of the DM halo during its past few revolutions around our Galaxy~\cite{SUBSTRUCTURE}. Time-dependent neutrino signals may also arise from the evolution of our Sun~\cite{Arellano:2021jul}, the Galactic supernova rate~\cite{Baum:2019fqm}, or the flux of cosmic rays hitting Earth's atmosphere~\cite{Jordan:2020gxx}. The code used in this work is available \href{\linkPaSens}{\texttt{here}}~\cite{PaleoSens}. Some of the options mentioned in this paragraph are already implemented in the code, and we invite the reader to add the capabilities in order to answer their own analysis questions.

\section{Summary and Conclusions} \label{sec:Conc}
This work surveys the prospects of searching for DM with paleo-detectors, natural minerals which can record and retain nuclear damage tracks for geological timescales. The exposure times of paleo-detectors can be of the order of a billion years, and modern microscopy techniques may allow for the readout of these damage tracks with nanometer spatial resolution in macroscopic samples. Thus, paleo-detectors could constitute nuclear recoil detectors with $\sim\,$keV recoil energy thresholds but with exposures rivaling those of large (planned) neutrino detectors such as Super/Hyper-Kamiokande or DUNE. Building on our previous work~\cite{Baum:2018tfw,Drukier:2018pdy,Edwards:2018hcf,Baum:2019fqm,Jordan:2020gxx}, we have discussed the relevant backgrounds for DM searches with paleo-detectors and presented updated sensitivity forecasts using a full spectral analysis. For the first time, the background induced by neutrinos from Galactic supernovae is included in our projections and, in this sense, the results presented here supersede the forecasts in Refs.~\cite{Baum:2018tfw,Drukier:2018pdy,Edwards:2018hcf}. We have also put particular emphasis on exploring the robustness of our sensitivity forecasts: we have demonstrated that DM-searches with paleo-detectors are possible even in the absence of any external information about the age of the samples, or the neutrino fluxes and the \Ur concentration, which control the normalization of the dominant backgrounds in paleo-detectors; the track length spectra alone contain sufficient information to infer these parameters. Furthermore, we have shown that the projected sensitivity is robust against uncertainties in the modeling of the spectral shape of the backgrounds. Finally, we have demonstrated that even if the uranium concentration in paleo-detector samples is $C^{238} = 10^{-8}\,$g/g, many orders of magnitude larger than what we expect in the most radiopure samples obtained from ultra basic rock or marine evaporite deposits, paleo-detectors could still probe WIMP-nucleon cross sections below current limits. In particular, for WIMP masses $m_\chi \lesssim 10\,$GeV/$c^2$, the sensitivity of paleo-detectors could still reach down all the way to the conventional neutrino floor in a Xe-based direct detection experiment. 

We hope that this article serves as further motivation for experimental efforts towards paleo-detectors. Ongoing efforts to demonstrate the feasibility of reading out tracks from keV-scale nuclear recoils~\cite{Hirose,RPI} are exciting first steps in this direction. Conventional direct detection experiments have been making impressive progress in their sensitivity to DM during the last few decades. However, despite detectors becoming ever larger and more sophisticated, they have failed to deliver (conclusive) evidence of WIMP-nucleon interactions to date. Future experiments are becoming increasingly expensive and challenging to operate. Thus, paleo-detectors are a timely proposal to probe the remaining WIMP parameter space. Moreover, paleo-detectors have applications to physics beyond DM searches, such as measuring solar, supernova, or atmospheric neutrinos, and offer a unique tool to explore the time-dependence of any signal over gigayear timescales. 

\acknowledgments
We thank our collaborators on previous and ongoing paleo-detector projects, Ethan~Brown, Francesco~Capozzi, William~DeRocco, Andrzej~Drukier, Alfredo~Ferrari, Maciej~G{\'o}rski, Shigenobu~Hirose, Shunsaku~Horiuchi, Jonathon~Jordan, Saarik~Kalia, Bradley~Kavanagh, Kirsten~McMichael, Maria~Morone, Kelly~Odgers, Paola~Sala, Morgan~Schaller, Joshua~Spitz, and Christoph~Weniger. We are deeply grateful to Frank~Avignone, John~Beacom, Juan~Collar, Rodney~Ewing, Ben~Feldman, Alfredo~Ferella, Ariel~Goobar, Peter~Graham, Chris~Kelso, Rafael~Lang, Dan~Snowden-Ifft, Kai~Sun and Martin~Winkler for discussion about various aspects of paleo-detectors.
The code used to produce the results in this work is available at: \href{\linkPaSpec}{\texttt{paleoSpec}}~\cite{PaleoSpec} for the computation of the signal and background spectra, and \href{\linkPaSens}{\texttt{paleoSens}}~\cite{PaleoSens} for the sensitivity forecast.
SB is supported in part by NSF Grant PHY-1720397, DOE HEP QuantISED award \#100495, and the Gordon and Betty Moore Foundation Grant GBMF7946. 
TE and KF acknowledge support by the Vetenskapsr{\aa}det (Swedish Research Council) through contract No.~638-2013-8993 and the Oskar Klein Centre for Cosmoparticle Physics.
KF is the Jeff \& Gail Kodosky Endowed Chair of Physics at the University of Texas in Austin and is grateful for support. 
PS is partially supported by the research grant ``The Dark Universe: A Synergetic Multi-messenger Approach'' number 2017X7X85K under the program PRIN~2017 funded by the Ministero dell'Istruzione, Universit{\`a} e della Ricerca (MIUR), and by the ``Hidden'' European ITN project (H2020-MSCA-ITN-2019//860881-HIDDeN).

\bibliographystyle{JHEP.bst}
\bibliography{theBib}

\providecommand{\href}[2]{#2}\begingroup\raggedright\begin{thebibliography}{100}

\bibitem{Bertone:2016nfn}
G.~Bertone and D.~Hooper, \emph{{History of dark matter}},
  \href{https://doi.org/10.1103/RevModPhys.90.045002}{\emph{Rev. Mod. Phys.}
  {\bfseries 90} (2018) 045002},
  [\href{https://arxiv.org/abs/1605.04909}{{\ttfamily 1605.04909}}].

\bibitem{Jungman:1995df}
G.~Jungman, M.~Kamionkowski and K.~Griest, \emph{{Supersymmetric dark matter}},
  \href{https://doi.org/10.1016/0370-1573(95)00058-5}{\emph{Phys. Rept.}
  {\bfseries 267} (1996) 195--373},
  [\href{https://arxiv.org/abs/hep-ph/9506380}{{\ttfamily hep-ph/9506380}}].

\bibitem{Arcadi:2017kky}
G.~Arcadi, M.~Dutra, P.~Ghosh, M.~Lindner, Y.~Mambrini, M.~Pierre et~al.,
  \emph{{The waning of the WIMP? A review of models, searches, and
  constraints}},
  \href{https://doi.org/10.1140/epjc/s10052-018-5662-y}{\emph{Eur. Phys. J. C}
  {\bfseries 78} (2018) 203},
  [\href{https://arxiv.org/abs/1703.07364}{{\ttfamily 1703.07364}}].

\bibitem{Roszkowski:2017nbc}
L.~Roszkowski, E.~M. Sessolo and S.~Trojanowski, \emph{{WIMP dark matter
  candidates and searches\textemdash{}current status and future prospects}},
  \href{https://doi.org/10.1088/1361-6633/aab913}{\emph{Rept. Prog. Phys.}
  {\bfseries 81} (2018) 066201},
  [\href{https://arxiv.org/abs/1707.06277}{{\ttfamily 1707.06277}}].

\bibitem{Drukier:1983gj}
A.~Drukier and L.~Stodolsky, \emph{{Principles and Applications of a Neutral
  Current Detector for Neutrino Physics and Astronomy}},
  \href{https://doi.org/10.1103/PhysRevD.30.2295}{\emph{Phys. Rev.} {\bfseries
  D30} (1984) 2295}.

\bibitem{Goodman:1984dc}
M.~W. Goodman and E.~Witten, \emph{{Detectability of Certain Dark Matter
  Candidates}}, \href{https://doi.org/10.1103/PhysRevD.31.3059}{\emph{Phys.
  Rev.} {\bfseries D31} (1985) 3059}.

\bibitem{Drukier:1986tm}
A.~K. Drukier, K.~Freese and D.~N. Spergel, \emph{{Detecting Cold Dark Matter
  Candidates}}, \href{https://doi.org/10.1103/PhysRevD.33.3495}{\emph{Phys.
  Rev.} {\bfseries D33} (1986) 3495--3508}.

\bibitem{Spergel:1987kx}
D.~N. Spergel, \emph{{The Motion of the Earth and the Detection of Wimps}},
  \href{https://doi.org/10.1103/PhysRevD.37.1353}{\emph{Phys. Rev.} {\bfseries
  D37} (1988) 1353}.

\bibitem{Collar:1992qc}
J.~I. Collar and F.~T. Avignone, \emph{{Diurnal modulation effects in cold dark
  matter experiments}},
  \href{https://doi.org/10.1016/0370-2693(92)90873-3}{\emph{Phys. Lett.}
  {\bfseries B275} (1992) 181--185}.

\bibitem{Akerib:2016vxi}
{\scshape LUX} collaboration, D.~S. Akerib et~al., \emph{{Results from a search
  for dark matter in the complete LUX exposure}},
  \href{https://doi.org/10.1103/PhysRevLett.118.021303}{\emph{Phys. Rev. Lett.}
  {\bfseries 118} (2017) 021303},
  [\href{https://arxiv.org/abs/1608.07648}{{\ttfamily 1608.07648}}].

\bibitem{Agnes:2018fwg}
{\scshape DarkSide} collaboration, P.~Agnes et~al., \emph{{DarkSide-50 532-day
  Dark Matter Search with Low-Radioactivity Argon}},
  \href{https://doi.org/10.1103/PhysRevD.98.102006}{\emph{Phys. Rev. D}
  {\bfseries 98} (2018) 102006},
  [\href{https://arxiv.org/abs/1802.07198}{{\ttfamily 1802.07198}}].

\bibitem{Aprile:2018dbl}
{\scshape XENON} collaboration, E.~Aprile et~al., \emph{{Dark Matter Search
  Results from a One Ton-Year Exposure of XENON1T}},
  \href{https://doi.org/10.1103/PhysRevLett.121.111302}{\emph{Phys. Rev. Lett.}
  {\bfseries 121} (2018) 111302},
  [\href{https://arxiv.org/abs/1805.12562}{{\ttfamily 1805.12562}}].

\bibitem{Wang:2020coa}
{\scshape PandaX-II} collaboration, Q.~Wang et~al., \emph{{Results of dark
  matter search using the full PandaX-II exposure}},
  \href{https://doi.org/10.1088/1674-1137/abb658}{\emph{Chin. Phys. C}
  {\bfseries 44} (2020) 125001},
  [\href{https://arxiv.org/abs/2007.15469}{{\ttfamily 2007.15469}}].

\bibitem{Angloher:2015ewa}
{\scshape CRESST} collaboration, G.~Angloher et~al., \emph{{Results on light
  dark matter particles with a low-threshold CRESST-II detector}},
  \href{https://doi.org/10.1140/epjc/s10052-016-3877-3}{\emph{Eur. Phys. J. C}
  {\bfseries 76} (2016) 25},
  [\href{https://arxiv.org/abs/1509.01515}{{\ttfamily 1509.01515}}].

\bibitem{Armengaud:2016cvl}
{\scshape EDELWEISS} collaboration, E.~Armengaud et~al., \emph{{Constraints on
  low-mass WIMPs from the EDELWEISS-III dark matter search}},
  \href{https://doi.org/10.1088/1475-7516/2016/05/019}{\emph{JCAP} {\bfseries
  05} (2016) 019}, [\href{https://arxiv.org/abs/1603.05120}{{\ttfamily
  1603.05120}}].

\bibitem{Agnese:2017jvy}
{\scshape SuperCDMS} collaboration, R.~Agnese et~al., \emph{{Low-mass dark
  matter search with CDMSlite}},
  \href{https://doi.org/10.1103/PhysRevD.97.022002}{\emph{Phys. Rev. D}
  {\bfseries 97} (2018) 022002},
  [\href{https://arxiv.org/abs/1707.01632}{{\ttfamily 1707.01632}}].

\bibitem{Petricca:2017zdp}
{\scshape CRESST} collaboration, F.~Petricca et~al., \emph{{First results on
  low-mass dark matter from the CRESST-III experiment}},
  \href{https://doi.org/10.1088/1742-6596/1342/1/012076}{\emph{J. Phys. Conf.
  Ser.} {\bfseries 1342} (2020) 012076},
  [\href{https://arxiv.org/abs/1711.07692}{{\ttfamily 1711.07692}}].

\bibitem{Armengaud:2019kfj}
{\scshape EDELWEISS} collaboration, E.~Armengaud et~al., \emph{{Searching for
  low-mass dark matter particles with a massive Ge bolometer operated
  above-ground}}, \href{https://doi.org/10.1103/PhysRevD.99.082003}{\emph{Phys.
  Rev. D} {\bfseries 99} (2019) 082003},
  [\href{https://arxiv.org/abs/1901.03588}{{\ttfamily 1901.03588}}].

\bibitem{Aprile:2015uzo}
{\scshape XENON} collaboration, E.~Aprile et~al., \emph{{Physics reach of the
  XENON1T dark matter experiment}},
  \href{https://doi.org/10.1088/1475-7516/2016/04/027}{\emph{JCAP} {\bfseries
  1604} (2016) 027}, [\href{https://arxiv.org/abs/1512.07501}{{\ttfamily
  1512.07501}}].

\bibitem{Mount:2017qzi}
B.~J. Mount et~al., \emph{{LUX-ZEPLIN (LZ) Technical Design Report}},
  \href{https://arxiv.org/abs/1703.09144}{{\ttfamily 1703.09144}}.

\bibitem{Aalbers:2016jon}
{\scshape DARWIN} collaboration, J.~Aalbers et~al., \emph{{DARWIN: towards the
  ultimate dark matter detector}},
  \href{https://doi.org/10.1088/1475-7516/2016/11/017}{\emph{JCAP} {\bfseries
  1611} (2016) 017}, [\href{https://arxiv.org/abs/1606.07001}{{\ttfamily
  1606.07001}}].

\bibitem{Aalseth:2017fik}
C.~E. Aalseth et~al., \emph{{DarkSide-20k: A 20 tonne two-phase LAr TPC for
  direct dark matter detection at LNGS}},
  \href{https://doi.org/10.1140/epjp/i2018-11973-4}{\emph{Eur. Phys. J. Plus}
  {\bfseries 133} (2018) 131},
  [\href{https://arxiv.org/abs/1707.08145}{{\ttfamily 1707.08145}}].

\bibitem{Amaudruz:2017ibl}
{\scshape DEAP-3600} collaboration, P.~A. Amaudruz et~al., \emph{{Design and
  Construction of the DEAP-3600 Dark Matter Detector}},
  \href{https://doi.org/10.1016/j.astropartphys.2018.09.006}{\emph{Astropart.
  Phys.} {\bfseries 108} (2019) 1--23},
  [\href{https://arxiv.org/abs/1712.01982}{{\ttfamily 1712.01982}}].

\bibitem{Daw:2011wq}
E.~Daw et~al., \emph{{The DRIFT Directional Dark Matter Experiments}},
  \href{https://doi.org/10.1051/eas/1253002}{\emph{EAS Publ. Ser.} {\bfseries
  53} (2012) 11--18}, [\href{https://arxiv.org/abs/1110.0222}{{\ttfamily
  1110.0222}}].

\bibitem{Cappella:2013rua}
F.~Cappella et~al., \emph{{On the potentiality of the $ZnWO_{4}$ anisotropic
  detectors to measure the directionality of Dark Matter}},
  \href{https://doi.org/10.1140/epjc/s10052-013-2276-2}{\emph{Eur. Phys. J. C}
  {\bfseries 73} (2013) 2276}.

\bibitem{Battat:2014van}
{\scshape DRIFT} collaboration, J.~B.~R. Battat et~al., \emph{{First
  background-free limit from a directional dark matter experiment: results from
  a fully fiducialised DRIFT detector}},
  \href{https://doi.org/10.1016/j.dark.2015.06.001}{\emph{Phys. Dark Univ.}
  {\bfseries 9-10} (2015) 1--7},
  [\href{https://arxiv.org/abs/1410.7821}{{\ttfamily 1410.7821}}].

\bibitem{Riffard:2013psa}
Q.~Riffard et~al., \emph{{Dark Matter directional detection with MIMAC}},  in
  \emph{{Proceedings, 48th Rencontres de Moriond on Very High Energy Phenomena
  in the Universe: La Thuile, Italy, March 9-16, 2013}}, pp.~227--230, 2013,
  \href{https://arxiv.org/abs/1306.4173}{{\ttfamily 1306.4173}},
  \href{http://inspirehep.net/record/1238915/files/arXiv:1306.4173.pdf}{http://inspirehep.net/record/1238915/files/arXiv:1306.4173.pdf}.

\bibitem{Santos:2013hpa}
D.~Santos et~al., \emph{{MIMAC: MIcro-tpc MAtrix of Chambers for dark matter
  directional detection}},
  \href{https://doi.org/10.1088/1742-6596/469/1/012002}{\emph{J. Phys. Conf.
  Ser.} {\bfseries 469} (2013) 012002},
  [\href{https://arxiv.org/abs/1311.0616}{{\ttfamily 1311.0616}}].

\bibitem{Monroe:2012qma}
{\scshape DMTPC} collaboration, J.~Monroe, \emph{{Status and Prospects of the
  DMTPC Directional Dark Matter Experiment}},
  \href{https://doi.org/10.1051/eas/1253003}{\emph{EAS Publ. Ser.} {\bfseries
  53} (2012) 19--24}.

\bibitem{Leyton:2016nit}
{\scshape DMTPC} collaboration, M.~Leyton, \emph{{Directional dark matter
  detection with the DMTPC m$^3$ experiment}},
  \href{https://doi.org/10.1088/1742-6596/718/4/042035}{\emph{J. Phys. Conf.
  Ser.} {\bfseries 718} (2016) 042035}.

\bibitem{Miuchi:2010hn}
K.~Miuchi et~al., \emph{{First underground results with NEWAGE-0.3a
  direction-sensitive dark matter detector}},
  \href{https://doi.org/10.1016/j.physletb.2010.02.028}{\emph{Phys. Lett.}
  {\bfseries B686} (2010) 11--17},
  [\href{https://arxiv.org/abs/1002.1794}{{\ttfamily 1002.1794}}].

\bibitem{Nakamura:2015iza}
K.~Nakamura et~al., \emph{{Direction-sensitive dark matter search with gaseous
  tracking detector NEWAGE-0.3b’}},
  \href{https://doi.org/10.1093/ptep/ptv041}{\emph{PTEP} {\bfseries 2015}
  (2015) 043F01}.

\bibitem{Battat:2016xxe}
{\scshape DRIFT} collaboration, J.~B.~R. Battat et~al., \emph{{Low Threshold
  Results and Limits from the DRIFT Directional Dark Matter Detector}},
  \href{https://doi.org/10.1016/j.astropartphys.2017.03.007}{\emph{Astropart.
  Phys.} {\bfseries 91} (2017) 65--74},
  [\href{https://arxiv.org/abs/1701.00171}{{\ttfamily 1701.00171}}].

\bibitem{Vahsen:2020pzb}
S.~E. Vahsen et~al., \emph{{CYGNUS: Feasibility of a nuclear recoil observatory
  with directional sensitivity to dark matter and neutrinos}},
  \href{https://arxiv.org/abs/2008.12587}{{\ttfamily 2008.12587}}.

\bibitem{Baum:2018tfw}
S.~Baum, A.~K. Drukier, K.~Freese, M.~G\'orski and P.~Stengel, \emph{{Searching
  for Dark Matter with Paleo-Detectors}},
  \href{https://doi.org/10.1016/j.physletb.2020.135325}{\emph{Phys. Lett. B}
  {\bfseries 803} (2020) 135325},
  [\href{https://arxiv.org/abs/1806.05991}{{\ttfamily 1806.05991}}].

\bibitem{Drukier:2018pdy}
A.~K. Drukier, S.~Baum, K.~Freese, M.~G\'orski and P.~Stengel,
  \emph{{Paleo-detectors: Searching for Dark Matter with Ancient Minerals}},
  \href{https://doi.org/10.1103/PhysRevD.99.043014}{\emph{Phys. Rev. D}
  {\bfseries 99} (2019) 043014},
  [\href{https://arxiv.org/abs/1811.06844}{{\ttfamily 1811.06844}}].

\bibitem{Edwards:2018hcf}
T.~D.~P. Edwards, B.~J. Kavanagh, C.~Weniger, S.~Baum, A.~K. Drukier, K.~Freese
  et~al., \emph{{Digging for dark matter: Spectral analysis and discovery
  potential of paleo-detectors}},
  \href{https://doi.org/10.1103/PhysRevD.99.043541}{\emph{Phys. Rev. D}
  {\bfseries 99} (2019) 043541},
  [\href{https://arxiv.org/abs/1811.10549}{{\ttfamily 1811.10549}}].

\bibitem{Fleischer:1964}
R.~L. Fleischer, P.~B. Price, R.~M. Walker and E.~L. Hubbard, \emph{Track
  registration in various solid-state nuclear track detectors},
  \href{https://doi.org/10.1103/PhysRev.133.A1443}{\emph{Phys. Rev.} {\bfseries
  133} (Mar, 1964) A1443--A1449}.

\bibitem{Fleischer383}
R.~L. Fleischer, P.~B. Price and R.~M. Walker, \emph{Tracks of charged
  particles in solids},
  \href{https://doi.org/10.1126/science.149.3682.383}{\emph{Science} {\bfseries
  149} (1965) 383--393}.

\bibitem{Fleischer:1965yv}
R.~L. Fleischer, P.~B. Price and R.~M. Walker, \emph{{Solid-state track
  detectors: applications to nuclear science and geophysics}},
  \href{https://doi.org/10.1146/annurev.ns.15.120165.000245}{\emph{Ann. Rev.
  Nucl. Part. Sci.} {\bfseries 15} (1965) 1--28}.

\bibitem{GUO2012233}
S.-L. Guo, B.-L. Chen and S.~Durrani, \emph{Chapter 4 - solid-state nuclear
  track detectors},  in \emph{Handbook of Radioactivity Analysis (Third
  Edition)} (M.~F. L'Annunziata, ed.), pp.~233 -- 298.
\newblock Academic Press, Amsterdam, third edition~ed., 2012.
\newblock \href{https://doi.org/10.1016/B978-0-12-384873-4.00004-9}{DOI}.

\bibitem{GTS2012}
M.~D.~S. Felix M.~Gradstein, James G.~Ogg and G.~M. Ogg, eds., \emph{The
  Geologic Time Scale}.
\newblock Elsevier, 2012,
  \href{https://doi.org/10.1016/c2011-1-08249-8}{10.1016/c2011-1-08249-8}.

\bibitem{Gallagher:1998}
K.~Gallagher, R.~Brown and C.~Johnson, \emph{Fission track analysis and its
  applications to geological problems},
  \href{https://doi.org/10.1146/annurev.earth.26.1.519}{\emph{Annual Review of
  Earth and Planetary Sciences} {\bfseries 26} (1998) 519--572},
  [\href{https://arxiv.org/abs/https://doi.org/10.1146/annurev.earth.26.1.519}{{\ttfamily
  https://doi.org/10.1146/annurev.earth.26.1.519}}].

\bibitem{vandenHaute:1998}
P.~van~den Haute and F.~de~Corte, eds., \emph{Advances in Fission-Track
  Geochronology}.
\newblock Springer, 1998,
  \href{https://doi.org/10.1007/978-94-015-9133-1}{10.1007/978-94-015-9133-1}.

\bibitem{Toulemonde:2006}
M.~Toulemonde, W.~Assmann, C.~Dufour, A.~Meftah, F.~Studer and C.~Trautmann,
  \emph{{Experimental Phenomena and Thermal Spike Model Description of Ion
  Tracks in Amorphisable Inorganic Insulators}},  in \emph{Ion Beam Science:
  Solved and Unsolved problems, Mat. Fys. Medd. Dan. Vid. Selsk.} (P.~Sigmund,
  ed.), vol.~52, (Copenhagen), p.~263, Det Kongelige Danske Videnskabernes
  Selskab, 2006.

\bibitem{HILL201265}
R.~Hill, J.~A. Notte and L.~Scipioni, \emph{Chapter 2 - scanning helium ion
  microscopy},  in \emph{Advances in Imaging and Electron Physics} (P.~W.
  Hawkes, ed.), vol.~170 of \emph{Advances in Imaging and Electron Physics},
  pp.~65 -- 148.
\newblock Elsevier, 2012.
\newblock \href{https://doi.org/10.1016/B978-0-12-394396-5.00002-6}{DOI}.

\bibitem{VANGASTEL20122104}
R.~van Gastel, G.~Hlawacek, H.~J. Zandvliet and B.~Poelsema, \emph{Subsurface
  analysis of semiconductor structures with helium ion microscopy},
  \href{https://doi.org/10.1016/j.microrel.2012.06.130}{\emph{Microelectronics
  Reliability} {\bfseries 52} (2012) 2104 -- 2109}.

\bibitem{RODRIGUEZ2014150}
M.~Rodriguez, W.~Li, F.~Chen, C.~Trautmann, T.~Bierschenk, B.~Afra et~al.,
  \emph{Saxs and tem investigation of ion tracks in neodymium-doped yttrium
  aluminium garnet},
  \href{https://doi.org/10.1016/j.nimb.2013.10.076}{\emph{Nuclear Instruments
  and Methods in Physics Research Section B: Beam Interactions with Materials
  and Atoms} {\bfseries 326} (2014) 150 -- 153}.

\bibitem{SAXS3d}
F.~Schaff, M.~Bech, P.~Zaslansky, C.~Jud, M.~Liebi, M.~Guizar-Sicairos et~al.,
  \emph{Six-dimensional real and reciprocal space small-angle x-ray scattering
  tomography}, \href{https://doi.org/10.1038/nature16060}{\emph{Nature}
  {\bfseries 527} (11, 2015) 353}.

\bibitem{SAXSres}
M.~Holler, A.~Diaz, M.~Guizar-Sicairos, P.~Karvinen, E.~F{\"a}rm,
  E.~H{\"a}rk{\"o}nen et~al., \emph{X-ray ptychographic computed tomography at
  16 nm isotropic 3d resolution},
  \href{https://doi.org/10.1038/srep03857}{\emph{Scientific Reports} {\bfseries
  4} (01, 2014) 3857}.

\bibitem{BARTZ2013273}
J.~Bartz, C.~Zeissler, V.~Fomenko and M.~Akselrod, \emph{An imaging
  spectrometer based on high resolution microscopy of fluorescent aluminum
  oxide crystal detectors},
  \href{https://doi.org/10.1016/j.radmeas.2013.01.041}{\emph{Radiation
  Measurements} {\bfseries 56} (2013) 273 -- 276}.

\bibitem{Kouwenberg:2018}
J.~Kouwenberg, G.~Kremers, J.~Slotman, H.~Woltenbeek, A.~Houtsmuller,
  A.~Denkova et~al., \emph{Alpha particle spectroscopy using fntd and sim
  super-resolution microscopy},
  \href{https://doi.org/10.1111/jmi.12686}{\emph{Journal of Microscopy}
  {\bfseries 270} 326--334}.

\bibitem{bioHIB}
M.~S. Joens, C.~Huynh, J.~M. Kasuboski, D.~Ferranti, Y.~J. Sigal, F.~Zeitvogel
  et~al., \emph{Helium ion microscopy (him) for the imaging of biological
  samples at sub-nanometer resolution},
  \href{https://doi.org/10.1038/srep03514}{\emph{Sci. Rep.} {\bfseries 3} (12,
  2013) 3514}.

\bibitem{ECHLIN20151}
M.~P. Echlin, M.~Straw, S.~Randolph, J.~Filevich and T.~M. Pollock, \emph{The
  tribeam system: Femtosecond laser ablation in situ sem},
  \href{https://doi.org/10.1016/j.matchar.2014.10.023}{\emph{Materials
  Characterization} {\bfseries 100} (2015) 1 -- 12}.

\bibitem{PFEIFENBERGER2017109}
M.~J. Pfeifenberger, M.~Mangang, S.~Wurster, J.~Reiser, A.~Hohenwarter,
  W.~Pfleging et~al., \emph{The use of femtosecond laser ablation as a novel
  tool for rapid micro-mechanical sample preparation},
  \href{https://doi.org/10.1016/j.matdes.2017.02.012}{\emph{Materials \&
  Design} {\bfseries 121} (2017) 109 -- 118}.

\bibitem{Randolph:2018}
S.~J. Randolph, J.~Filevich, A.~Botman, R.~Gannon, C.~Rue and M.~Straw,
  \emph{{In situ femtosecond pulse laser ablation for large volume 3D analysis
  in scanning electron microscope systems}},
  \href{https://doi.org/10.1116/1.5047806}{\emph{Journal of Vacuum Science \&
  Technology B} {\bfseries 36} (2018) 06JB01}.

\bibitem{Goto:1958}
E.~Goto, \emph{{On the Observation of Magnetic Monopoles}},
  \href{https://doi.org/10.1143/JPSJ.13.1413}{\emph{J. Phys. Soc. Japan}
  {\bfseries 13} (1958) 1413--1418}.

\bibitem{Goto:1963zz}
E.~Goto, H.~H. Kolm and K.~W. Ford, \emph{{Search for Ferromagnetically Trapped
  Magnetic Monopoles of Cosmic-Ray Origin}},
  \href{https://doi.org/10.1103/PhysRev.132.387}{\emph{Phys. Rev.} {\bfseries
  132} (1963) 387}.

\bibitem{Fleischer:1969mj}
R.~L. Fleischer, I.~S. Jacobs, W.~M. Schwarz, P.~B. Price and H.~G. Goodell,
  \emph{{Search for multiply charged dirac magnetic poles}},
  \href{https://doi.org/10.1103/PhysRev.177.2029}{\emph{Phys. Rev.} {\bfseries
  177} (1969) 2029--2035}.

\bibitem{Fleischer:1970zy}
R.~L. Fleischer, H.~R. Hart, I.~S. Jacobs, P.~B. Price, W.~M. Schwarz and
  F.~Aumento, \emph{{Search for magnetic monopoles in deep ocean deposits}},
  \href{https://doi.org/10.1103/PhysRev.184.1393}{\emph{Phys. Rev.} {\bfseries
  184} (1969) 1393--1397}.

\bibitem{Fleischer:1970vm}
R.~L. Fleischer, P.~B. Price and R.~T. Woods, \emph{{Search for tracks of
  massive, multiply charged magnetic poles}},
  \href{https://doi.org/10.1103/PhysRev.184.1398}{\emph{Phys. Rev.} {\bfseries
  184} (1969) 1398--1401}.

\bibitem{Alvarez:1970zu}
L.~W. Alvarez, P.~H. Eberhard, R.~R. Ross and R.~D. Watt, \emph{{Search for
  Magnetic Monopoles in the Lunar Sample}},
  \href{https://doi.org/10.1126/science.167.3918.701}{\emph{Science} {\bfseries
  167} (1970) 701--703}.

\bibitem{Kolm:1971xb}
H.~H. Kolm, F.~Villa and A.~Odian, \emph{{SEARCH FOR MAGNETIC MONOPOLES}},
  \href{https://doi.org/10.1103/PhysRevD.4.1285}{\emph{Phys. Rev.} {\bfseries
  D4} (1971) 1285}.

\bibitem{Eberhard:1971re}
P.~H. Eberhard, R.~R. Ross, L.~W. Alvarez and R.~D. Watt, \emph{{SEARCH FOR
  MAGNETIC MONOPOLES IN LUNAR MATERIAL}},
  \href{https://doi.org/10.1103/PhysRevD.4.3260}{\emph{Phys. Rev.} {\bfseries
  D4} (1971) 3260}.

\bibitem{Ross:1973it}
R.~R. Ross, P.~H. Eberhard, L.~W. Alvarez and R.~D. Watt, \emph{{SEARCH FOR
  MAGNETIC MONOPOLES IN LUNAR MATERIAL USING AN ELECTROMAGNETIC DETECTOR}},
  \href{https://doi.org/10.1103/PhysRevD.8.698}{\emph{Phys. Rev.} {\bfseries
  D8} (1973) 698}.

\bibitem{Price:1983ax}
P.~B. Price, S.-l. Guo, S.~P. Ahlen and R.~L. Fleischer, \emph{{Search for
  {GUT} Magnetic Monopoles at a Flux Level Below the Parker Limit}},
  \href{https://doi.org/10.1103/PhysRevLett.52.1265}{\emph{Phys. Rev. Lett.}
  {\bfseries 52} (1984) 1265}.

\bibitem{Kovalik:1986zz}
J.~M. Kovalik and J.~L. Kirschvink, \emph{{New Superconducting Quantum
  Interface Device Based Constraints on the Abundance of Magnetic Monopoles
  Trapped in Matter: An Investigation of Deeply Buried Rocks}},
  \href{https://doi.org/10.1103/PhysRevA.33.1183}{\emph{Phys. Rev.} {\bfseries
  A33} (1986) 1183--1187}.

\bibitem{Price:1986ky}
P.~B. Price and M.~H. Salamon, \emph{{Search for Supermassive Magnetic
  Monopoles Using Mica Crystals}},
  \href{https://doi.org/10.1103/PhysRevLett.56.1226}{\emph{Phys. Rev. Lett.}
  {\bfseries 56} (1986) 1226--1229}.

\bibitem{Ghosh:1990ki}
D.~Ghosh and S.~Chatterjea, \emph{{Supermassive magnetic monopoles flux from
  the oldest mica samples}},
  \href{https://doi.org/10.1209/0295-5075/12/1/005}{\emph{Europhys. Lett.}
  {\bfseries 12} (1990) 25--28}.

\bibitem{Jeon:1995rf}
H.~Jeon and M.~J. Longo, \emph{{Search for magnetic monopoles trapped in
  matter}}, \href{https://doi.org/10.1103/PhysRevLett.75.1443}{\emph{Phys. Rev.
  Lett.} {\bfseries 75} (1995) 1443--1446},
  [\href{https://arxiv.org/abs/hep-ex/9508003}{{\ttfamily hep-ex/9508003}}].

\bibitem{Collar:1999md}
J.~I. Collar and K.~Zioutas, \emph{{Limits on exotic heavily ionizing particles
  from the geological abundance of fullerenes}},
  \href{https://doi.org/10.1103/PhysRevLett.83.3097}{\emph{Phys. Rev. Lett.}
  {\bfseries 83} (1999) 3097},
  [\href{https://arxiv.org/abs/astro-ph/9902310}{{\ttfamily
  astro-ph/9902310}}].

\bibitem{SnowdenIfft:1995ke}
D.~P. Snowden-Ifft, E.~S. Freeman and P.~B. Price, \emph{{Limits on dark matter
  using ancient mica}},
  \href{https://doi.org/10.1103/PhysRevLett.74.4133}{\emph{Phys. Rev. Lett.}
  {\bfseries 74} (1995) 4133--4136}.

\bibitem{Collar:1994mj}
J.~I. Collar and F.~T. Avignone, III, \emph{{Nuclear tracks from cold dark
  matter interactions in mineral crystals: A Computational study}},
  \href{https://doi.org/10.1016/0168-583X(94)00543-5}{\emph{Nucl. Instrum.
  Meth.} {\bfseries B95} (1995) 349},
  [\href{https://arxiv.org/abs/astro-ph/9505055}{{\ttfamily
  astro-ph/9505055}}].

\bibitem{Engel:1995gw}
J.~Engel, M.~T. Ressell, I.~S. Towner and W.~E. Ormand, \emph{{Response of mica
  to weakly interacting massive particles}},
  \href{https://doi.org/10.1103/PhysRevC.52.2216}{\emph{Phys. Rev.} {\bfseries
  C52} (1995) 2216--2221},
  [\href{https://arxiv.org/abs/hep-ph/9504322}{{\ttfamily hep-ph/9504322}}].

\bibitem{SnowdenIfft:1997hd}
D.~P. Snowden-Ifft and A.~J. Westphal, \emph{{Unique signature of dark matter
  in ancient mica}},
  \href{https://doi.org/10.1103/PhysRevLett.78.1628}{\emph{Phys. Rev. Lett.}
  {\bfseries 78} (1997) 1628--1631},
  [\href{https://arxiv.org/abs/astro-ph/9701215}{{\ttfamily
  astro-ph/9701215}}].

\bibitem{Baum:2019fqm}
S.~Baum, T.~D.~P. Edwards, B.~J. Kavanagh, P.~Stengel, A.~K. Drukier, K.~Freese
  et~al., \emph{{Paleodetectors for Galactic supernova neutrinos}},
  \href{https://doi.org/10.1103/PhysRevD.101.103017}{\emph{Phys. Rev. D}
  {\bfseries 101} (2020) 103017},
  [\href{https://arxiv.org/abs/1906.05800}{{\ttfamily 1906.05800}}].

\bibitem{Jordan:2020gxx}
J.~R. Jordan, S.~Baum, P.~Stengel, A.~Ferrari, M.~C. Morone, P.~Sala et~al.,
  \emph{{Measuring Changes in the Atmospheric Neutrino Rate Over Gigayear
  Timescales}},
  \href{https://doi.org/10.1103/PhysRevLett.125.231802}{\emph{Phys. Rev. Lett.}
  {\bfseries 125} (2020) 231802},
  [\href{https://arxiv.org/abs/2004.08394}{{\ttfamily 2004.08394}}].

\bibitem{Arellano:2021jul}
N.~T. Arellano and S.~Horiuchi, \emph{{Measuring solar neutrinos over Gigayear
  timescales with Paleo Detectors}},
  \href{https://arxiv.org/abs/2102.01755}{{\ttfamily 2102.01755}}.

\bibitem{Rajendran:2017ynw}
S.~Rajendran, N.~Zobrist, A.~O. Sushkov, R.~Walsworth and M.~Lukin, \emph{{A
  method for directional detection of dark matter using spectroscopy of crystal
  defects}}, \href{https://doi.org/10.1103/PhysRevD.96.035009}{\emph{Phys. Rev.
  D} {\bfseries 96} (2017) 035009},
  [\href{https://arxiv.org/abs/1705.09760}{{\ttfamily 1705.09760}}].

\bibitem{Essig:2016crl}
R.~Essig, J.~Mardon, O.~Slone and T.~Volansky, \emph{{Detection of sub-GeV Dark
  Matter and Solar Neutrinos via Chemical-Bond Breaking}},
  \href{https://doi.org/10.1103/PhysRevD.95.056011}{\emph{Phys. Rev. D}
  {\bfseries 95} (2017) 056011},
  [\href{https://arxiv.org/abs/1608.02940}{{\ttfamily 1608.02940}}].

\bibitem{Budnik:2017sbu}
R.~Budnik, O.~Chesnovsky, O.~Slone and T.~Volansky, \emph{{Direct Detection of
  Light Dark Matter and Solar Neutrinos via Color Center Production in
  Crystals}}, \href{https://doi.org/10.1016/j.physletb.2018.04.063}{\emph{Phys.
  Lett. B} {\bfseries 782} (2018) 242--250},
  [\href{https://arxiv.org/abs/1705.03016}{{\ttfamily 1705.03016}}].

\bibitem{Cogswell:2021qlq}
B.~K. Cogswell, A.~Goel and P.~Huber, \emph{{Passive low-energy nuclear recoil
  detection with color centers}},
  \href{https://arxiv.org/abs/2104.13926}{{\ttfamily 2104.13926}}.

\bibitem{Ebadi:2021cte}
R.~Ebadi et~al., \emph{{Ultra-Heavy Dark Matter Search with Electron Microscopy
  of Geological Quartz}},  \href{https://arxiv.org/abs/2105.03998}{{\ttfamily
  2105.03998}}.

\bibitem{SUBSTRUCTURE}
S.~Baum, W.~DeRocco, T.~D.~P. Edwards and S.~Kalia, \emph{Throwing rocks at
  dinosaurs: Time-varying dark matter signals in paleo-detectors}, .

\bibitem{PaleoSpec}
\url{https://github.com/sbaum90/paleoSpec}.

\bibitem{PaleoSens}
\url{https://github.com/sbaum90/paleoSens}.

\bibitem{Engel:1991wq}
J.~Engel, \emph{{Nuclear form-factors for the scattering of weakly interacting
  massive particles}},
  \href{https://doi.org/10.1016/0370-2693(91)90712-Y}{\emph{Phys. Lett.}
  {\bfseries B264} (1991) 114--119}.

\bibitem{Engel:1992bf}
J.~Engel, S.~Pittel and P.~Vogel, \emph{{Nuclear physics of dark matter
  detection}}, \href{https://doi.org/10.1142/S0218301392000023}{\emph{Int. J.
  Mod. Phys.} {\bfseries E1} (1992) 1--37}.

\bibitem{Ressell:1993qm}
M.~T. Ressell, M.~B. Aufderheide, S.~D. Bloom, K.~Griest, G.~J. Mathews and
  D.~A. Resler, \emph{{Nuclear shell model calculations of neutralino - nucleus
  cross-sections for Si-29 and Ge-73}},
  \href{https://doi.org/10.1103/PhysRevD.48.5519}{\emph{Phys. Rev.} {\bfseries
  D48} (1993) 5519--5535},
  [\href{https://arxiv.org/abs/hep-ph/9307228}{{\ttfamily hep-ph/9307228}}].

\bibitem{Bednyakov:2004xq}
V.~A. Bednyakov and F.~Simkovic, \emph{{Nuclear spin structure in dark matter
  search: The Zero momentum transfer limit}}, {\emph{Phys. Part. Nucl.}
  {\bfseries 36} (2005) 131--152},
  [\href{https://arxiv.org/abs/hep-ph/0406218}{{\ttfamily hep-ph/0406218}}].

\bibitem{Bednyakov:2006ux}
V.~A. Bednyakov and F.~Simkovic, \emph{{Nuclear spin structure in dark matter
  search: The Finite momentum transfer limit}},
  \href{https://doi.org/10.1134/S1063779606070057}{\emph{Phys. Part. Nucl.}
  {\bfseries 37} (2006) S106--S128},
  [\href{https://arxiv.org/abs/hep-ph/0608097}{{\ttfamily hep-ph/0608097}}].

\bibitem{Fan:2010gt}
J.~Fan, M.~Reece and L.-T. Wang, \emph{{Non-relativistic effective theory of
  dark matter direct detection}},
  \href{https://doi.org/10.1088/1475-7516/2010/11/042}{\emph{JCAP} {\bfseries
  1011} (2010) 042}, [\href{https://arxiv.org/abs/1008.1591}{{\ttfamily
  1008.1591}}].

\bibitem{Fitzpatrick:2012ix}
A.~L. Fitzpatrick, W.~Haxton, E.~Katz, N.~Lubbers and Y.~Xu, \emph{{The
  Effective Field Theory of Dark Matter Direct Detection}},
  \href{https://doi.org/10.1088/1475-7516/2013/02/004}{\emph{JCAP} {\bfseries
  1302} (2013) 004}, [\href{https://arxiv.org/abs/1203.3542}{{\ttfamily
  1203.3542}}].

\bibitem{Helm:1956zz}
R.~H. Helm, \emph{{Inelastic and Elastic Scattering of 187-Mev Electrons from
  Selected Even-Even Nuclei}},
  \href{https://doi.org/10.1103/PhysRev.104.1466}{\emph{Phys. Rev.} {\bfseries
  104} (1956) 1466--1475}.

\bibitem{Lewin:1995rx}
J.~D. Lewin and P.~F. Smith, \emph{{Review of mathematics, numerical factors,
  and corrections for dark matter experiments based on elastic nuclear
  recoil}},
  \href{https://doi.org/10.1016/S0927-6505(96)00047-3}{\emph{Astropart. Phys.}
  {\bfseries 6} (1996) 87--112}.

\bibitem{Duda:2006uk}
G.~Duda, A.~Kemper and P.~Gondolo, \emph{{Model Independent Form Factors for
  Spin Independent Neutralino-Nucleon Scattering from Elastic Electron
  Scattering Data}},
  \href{https://doi.org/10.1088/1475-7516/2007/04/012}{\emph{JCAP} {\bfseries
  0704} (2007) 012}, [\href{https://arxiv.org/abs/hep-ph/0608035}{{\ttfamily
  hep-ph/0608035}}].

\bibitem{Vietze:2014vsa}
L.~Vietze, P.~Klos, J.~Menéndez, W.~C. Haxton and A.~Schwenk, \emph{{Nuclear
  structure aspects of spin-independent WIMP scattering off xenon}},
  \href{https://doi.org/10.1103/PhysRevD.91.043520}{\emph{Phys. Rev.}
  {\bfseries D91} (2015) 043520},
  [\href{https://arxiv.org/abs/1412.6091}{{\ttfamily 1412.6091}}].

\bibitem{Gazda:2016mrp}
D.~Gazda, R.~Catena and C.~Forssén, \emph{{Ab initio nuclear response
  functions for dark matter searches}},
  \href{https://doi.org/10.1103/PhysRevD.95.103011}{\emph{Phys. Rev.}
  {\bfseries D95} (2017) 103011},
  [\href{https://arxiv.org/abs/1612.09165}{{\ttfamily 1612.09165}}].

\bibitem{Korber:2017ery}
C.~Körber, A.~Nogga and J.~de~Vries, \emph{{First-principle calculations of
  Dark Matter scattering off light nuclei}},
  \href{https://doi.org/10.1103/PhysRevC.96.035805}{\emph{Phys. Rev.}
  {\bfseries C96} (2017) 035805},
  [\href{https://arxiv.org/abs/1704.01150}{{\ttfamily 1704.01150}}].

\bibitem{Hoferichter:2018acd}
M.~Hoferichter, P.~Klos, J.~Men\'endez and A.~Schwenk, \emph{{Nuclear structure
  factors for general spin-independent WIMP-nucleus scattering}},
  \href{https://doi.org/10.1103/PhysRevD.99.055031}{\emph{Phys. Rev. D}
  {\bfseries 99} (2019) 055031},
  [\href{https://arxiv.org/abs/1812.05617}{{\ttfamily 1812.05617}}].

\bibitem{Koposov:2009hn}
S.~E. Koposov, H.-W. Rix and D.~W. Hogg, \emph{{Constraining the Milky Way
  potential with a 6-D phase-space map of the GD-1 stellar stream}},
  \href{https://doi.org/10.1088/0004-637X/712/1/260}{\emph{Astrophys. J.}
  {\bfseries 712} (2010) 260--273},
  [\href{https://arxiv.org/abs/0907.1085}{{\ttfamily 0907.1085}}].

\bibitem{Piffl:2013mla}
T.~Piffl et~al., \emph{{The RAVE survey: the Galactic escape speed and the mass
  of the Milky Way}},
  \href{https://doi.org/10.1051/0004-6361/201322531}{\emph{Astron. Astrophys.}
  {\bfseries 562} (2014) A91},
  [\href{https://arxiv.org/abs/1309.4293}{{\ttfamily 1309.4293}}].

\bibitem{Bovy:2012ba}
J.~Bovy et~al., \emph{{The Milky Way's circular velocity curve between 4 and 14
  kpc from APOGEE data}},
  \href{https://doi.org/10.1088/0004-637X/759/2/131}{\emph{Astrophys. J.}
  {\bfseries 759} (2012) 131},
  [\href{https://arxiv.org/abs/1209.0759}{{\ttfamily 1209.0759}}].

\bibitem{Freese:2012xd}
K.~Freese, M.~Lisanti and C.~Savage, \emph{{Colloquium: Annual modulation of
  dark matter}}, \href{https://doi.org/10.1103/RevModPhys.85.1561}{\emph{Rev.
  Mod. Phys.} {\bfseries 85} (2013) 1561--1581},
  [\href{https://arxiv.org/abs/1209.3339}{{\ttfamily 1209.3339}}].

\bibitem{Aprile:2019xxb}
{\scshape XENON} collaboration, E.~Aprile et~al., \emph{{Light Dark Matter
  Search with Ionization Signals in XENON1T}},
  \href{https://doi.org/10.1103/PhysRevLett.123.251801}{\emph{Phys. Rev. Lett.}
  {\bfseries 123} (2019) 251801},
  [\href{https://arxiv.org/abs/1907.11485}{{\ttfamily 1907.11485}}].

\bibitem{Ziegler:1985}
J.~F. Ziegler, J.~P. Biersack and U.~Littmark, \emph{{The Stopping and Range of
  Ions in Solids}}.
\newblock Pergamon Press, New York, 01, 1985.

\bibitem{Ziegler:2010}
J.~F. Ziegler, M.~D. Ziegler and J.~P. Biersack, \emph{{SRIM – The stopping
  and range of ions in matter (2010)}},
  \href{https://doi.org/10.1016/j.nimb.2010.02.091}{\emph{Nuclear Instruments
  and Methods in Physics Research Section B: Beam Interactions with Materials
  and Atoms} {\bfseries 268} (2010) 1818 -- 1823}.

\bibitem{Hirose}
S.~Hirose. private communication.

\bibitem{RPI}
E.~Brown. private communication.

\bibitem{Mei:2005gm}
D.~Mei and A.~Hime, \emph{{Muon-induced background study for underground
  laboratories}}, \href{https://doi.org/10.1103/PhysRevD.73.053004}{\emph{Phys.
  Rev.} {\bfseries D73} (2006) 053004},
  [\href{https://arxiv.org/abs/astro-ph/0512125}{{\ttfamily
  astro-ph/0512125}}].

\bibitem{Aharmim:2009zm}
{\scshape SNO} collaboration, B.~Aharmim et~al., \emph{{Measurement of the
  Cosmic Ray and Neutrino-Induced Muon Flux at the Sudbury Neutrino
  Observatory}}, \href{https://doi.org/10.1103/PhysRevD.80.012001}{\emph{Phys.
  Rev. D} {\bfseries 80} (2009) 012001},
  [\href{https://arxiv.org/abs/0902.2776}{{\ttfamily 0902.2776}}].

\bibitem{OHare:2020lva}
C.~A.~J. O'Hare, \emph{{Can we overcome the neutrino floor at high masses?}},
  \href{https://doi.org/10.1103/PhysRevD.102.063024}{\emph{Phys. Rev. D}
  {\bfseries 102} (2020) 063024},
  [\href{https://arxiv.org/abs/2002.07499}{{\ttfamily 2002.07499}}].

\bibitem{Cappellaro:2003eg}
E.~Cappellaro, R.~Barbon and M.~Turatto, \emph{{Supernova statistics}},
  \href{https://doi.org/10.1007/3-540-26633-X_48}{\emph{Springer Proc. Phys.}
  {\bfseries 99} (2005) 347--354},
  [\href{https://arxiv.org/abs/astro-ph/0310859}{{\ttfamily
  astro-ph/0310859}}].

\bibitem{Diehl:2006cf}
R.~Diehl et~al., \emph{{Radioactive Al-26 and massive stars in the galaxy}},
  \href{https://doi.org/10.1038/nature04364}{\emph{Nature} {\bfseries 439}
  (2006) 45--47}, [\href{https://arxiv.org/abs/astro-ph/0601015}{{\ttfamily
  astro-ph/0601015}}].

\bibitem{Strumia:2006db}
A.~Strumia and F.~Vissani, \emph{{Neutrino masses and mixings and...}},
  \href{https://arxiv.org/abs/hep-ph/0606054}{{\ttfamily hep-ph/0606054}}.

\bibitem{Li:2011}
W.~Li, R.~Chornock, J.~Leaman, A.~V. Filippenko, D.~Poznanski, X.~Wang et~al.,
  \emph{{Nearby supernova rates from the Lick Observatory Supernova Search -
  III. The rate-size relation, and the rates as a function of galaxy Hubble
  type and colour}},
  \href{https://doi.org/10.1111/j.1365-2966.2011.18162.x}{\emph{Mon. Not. Roy.
  Astron. Soc.} {\bfseries 412} (2011) 1473--1507},
  [\href{https://arxiv.org/abs/1006.4613}{{\ttfamily 1006.4613}}].

\bibitem{Botticella:2011nd}
M.~T. Botticella, S.~J. Smartt, R.~C. Kennicutt, Jr., E.~Cappellaro, M.~Sereno
  and J.~C. Lee, \emph{{A comparison between star formation rate diagnostics
  and rate of core collapse supernovae within 11 Mpc}},
  \href{https://doi.org/10.1051/0004-6361/201117343}{\emph{Astron. Astrophys.}
  {\bfseries 537} (2012) A132},
  [\href{https://arxiv.org/abs/1111.1692}{{\ttfamily 1111.1692}}].

\bibitem{Adams:2013ana}
S.~M. Adams, C.~S. Kochanek, J.~F. Beacom, M.~R. Vagins and K.~Z. Stanek,
  \emph{{Observing the Next Galactic Supernova}},
  \href{https://doi.org/10.1088/0004-637X/778/2/164}{\emph{Astrophys. J.}
  {\bfseries 778} (2013) 164},
  [\href{https://arxiv.org/abs/1306.0559}{{\ttfamily 1306.0559}}].

\bibitem{Beacom:2010kk}
J.~F. Beacom, \emph{{The Diffuse Supernova Neutrino Background}},
  \href{https://doi.org/10.1146/annurev.nucl.010909.083331}{\emph{Ann. Rev.
  Nucl. Part. Sci.} {\bfseries 60} (2010) 439--462},
  [\href{https://arxiv.org/abs/1004.3311}{{\ttfamily 1004.3311}}].

\bibitem{Billard:2013qya}
J.~Billard, L.~Strigari and E.~Figueroa-Feliciano, \emph{{Implication of
  neutrino backgrounds on the reach of next generation dark matter direct
  detection experiments}},
  \href{https://doi.org/10.1103/PhysRevD.89.023524}{\emph{Phys. Rev.}
  {\bfseries D89} (2014) 023524},
  [\href{https://arxiv.org/abs/1307.5458}{{\ttfamily 1307.5458}}].

\bibitem{Madau:2014bja}
P.~Madau and M.~Dickinson, \emph{{Cosmic Star Formation History}},
  \href{https://doi.org/10.1146/annurev-astro-081811-125615}{\emph{Ann. Rev.
  Astron. Astrophys.} {\bfseries 52} (2014) 415--486},
  [\href{https://arxiv.org/abs/1403.0007}{{\ttfamily 1403.0007}}].

\bibitem{Strolger:2015kra}
L.-G. Strolger, T.~Dahlen, S.~A. Rodney, O.~Graur, A.~G. Riess, C.~McCully
  et~al., \emph{{The Rate of Core Collapse Supernovae to Redshift 2.5 From The
  CANDELS and CLASH Supernova Surveys}},
  \href{https://doi.org/10.1088/0004-637X/813/2/93}{\emph{Astrophys. J.}
  {\bfseries 813} (2015) 93},
  [\href{https://arxiv.org/abs/1509.06574}{{\ttfamily 1509.06574}}].

\bibitem{Thomson:1954}
S.~Thomson and G.~Wardle, \emph{{Coloured natural rocksalts: a study of their
  helium contents, colours and impurities}},
  \href{https://doi.org/10.1016/0016-7037(54)90031-9}{\emph{Geochimica et
  Cosmochimica Acta} {\bfseries 5} (1954) 169 -- 184}.

\bibitem{Condie:1957}
K.~C. Condie, C.~S. Kuo, R.~M. Walker and V.~R. Murthy, \emph{{Uranium
  Distribution in Separated Clinopyroxenes from Four Eclogites}},
  \href{https://doi.org/10.1126/science.165.3888.57}{\emph{Science} {\bfseries
  165} (1969) 57--59}.

\bibitem{Adams:1959}
J.~A.~S. Adams, J.~K. Osmond and J.~J.~W. Rogers, \emph{{The geochemistry of
  thorium and uranium}},
  \href{https://doi.org/10.1016/0079-1946(59)90008-4}{\emph{Phys. Chem. Earth}
  {\bfseries 3} (1959) 298--348}.

\bibitem{Seitz:1974}
M.~Seitz and S.~Hart, \emph{{Uranium and boron distributions in some oceanic
  ultramafic rocks}},
  \href{https://doi.org/10.1016/0012-821X(73)90230-6}{\emph{Earth and Planetary
  Science Letters} {\bfseries 21} (1973) 97 -- 107}.

\bibitem{Dean:1978}
W.~E. Dean, \emph{Section 5 trace and minor elements in evaporites},  in
  \emph{{Marine Evaporites}}.
\newblock SEPM Society for Sedimentary Geology, 1987.
\newblock \href{https://doi.org/10.2110/scn.78.01.0086}{DOI}.

\bibitem{Yui:1998}
M.~Yui, Y.~Kikawada, T.~Oi, T.~Honda and T.~Nozaki, \emph{{Abundance of Uranium
  and Thorium in Rock Salts}},
  \href{https://doi.org/10.3769/radioisotopes.47.488}{\emph{Radioisotopes}
  {\bfseries 47} (1998) 488--492}.

\bibitem{Sanford:2013}
W.~Sanford, M.~Doughten, T.~Coplen, A.~Hunt and T.~Bullen, \emph{{Evidence for
  high salinity of Early Cretaceous sea water from the Chesapeake Bay crater}},
  \href{https://doi.org/10.1038/nature12714}{\emph{Nature} {\bfseries 503}
  (2013) 252--256}.

\bibitem{Collar:1995aw}
J.~I. Collar, \emph{{Comments on 'limits on dark matter using ancient mica'}},
  \href{https://doi.org/10.1103/PhysRevLett.76.331}{\emph{Phys. Rev. Lett.}
  {\bfseries 76} (1996) 331},
  [\href{https://arxiv.org/abs/astro-ph/9511055}{{\ttfamily
  astro-ph/9511055}}].

\bibitem{SnowdenIfft:1996zz}
D.~P. Snowden-Ifft, E.~S. Freeman and P.~B. Price, \emph{{Snowden-Ifft,
  Freemen, and Price Reply (A Reply to the Comment by Juan I Collar)}},
  \href{https://doi.org/10.1103/PhysRevLett.76.332}{\emph{Phys. Rev. Lett.}
  {\bfseries 76} (1996) 332}.

\bibitem{sources4a:1999}
\emph{{SOURCES 4A: A Code for Calculating (alpha,n), Spontaneous Fission, and
  Delayed Neutron Sources and Spectra}},  Tech. Rep. LA-13639-MS, Los Alamos
  National Lab, 1999.
\newblock 10.2172/15215.

\bibitem{Koning:2012zqy}
A.~J. Koning and D.~Rochman, \emph{{Modern Nuclear Data Evaluation with the
  TALYS Code System}},
  \href{https://doi.org/10.1016/j.nds.2012.11.002}{\emph{Nucl. Data Sheets}
  {\bfseries 113} (2012) 2841--2934}.

\bibitem{Rochman:2016}
D.~Rochman, A.~J. Koning, J.~C. Sublet, M.~Fleming et~al., \emph{{The TENDL
  library: hope, reality and future}},  in \emph{{Proceedings of the
  International Conference on Nuclear Data for Science and Technology,
  September 11-16, 2016, Bruges, Belgium}}, 2016,
  \href{https://tendl.web.psi.ch/bib\_rochman/tendl.nd2016.pdf}{https://tendl.web.psi.ch/bib\_rochman/tendl.nd2016.pdf}.

\bibitem{Sublet:2015}
J.~C. Sublet, A.~J. Koning, D.~Rochman, M.~Fleming and M.~Gilbert,
  \emph{{TENDL-2015: Delivering Both Completeness and Robustness}},  in
  \emph{{Advances in Nuclear Nonproliferation Technology and Policy Conference,
  Sept. 25-30, Santa Fe, NM, USA}},
  \href{https://tendl.web.psi.ch/bib\_rochman/ANTPC\_TENDL.pdf}{https://tendl.web.psi.ch/bib\_rochman/ANTPC\_TENDL.pdf}.

\bibitem{Fleming:2015}
M.~Fleming, J.~C. Sublet, J.~Kopecky, D.~Rochman and A.~J. Koning,
  \emph{{Probing experimental and systematic trends of the neutron-induced
  TENDL-2014 nuclear data library}},  in \emph{{CCFE report UKAEA-R(15)29,
  October 2015}}, 2015,
  \href{https://tendl.web.psi.ch/bib\_rochman/UKAEA-R(15)29\_30.9.15.pdf}{https://tendl.web.psi.ch/bib\_rochman/UKAEA-R(15)29\_30.9.15.pdf}.

\bibitem{Soppera:2014zsj}
N.~Soppera, M.~Bossant and E.~Dupont, \emph{{JANIS 4: An Improved Version of
  the NEA Java-based Nuclear Data Information System}},
  \href{https://doi.org/10.1016/j.nds.2014.07.071}{\emph{Nucl. Data Sheets}
  {\bfseries 120} (2014) 294--296}.

\bibitem{Ferrari:2005zk}
A.~Ferrari, P.~R. Sala, A.~Fasso and J.~Ranft, \emph{{FLUKA: A multi-particle
  transport code (Program version 2005)}}, .

\bibitem{Bohlen:2014buj}
T.~T. Böhlen, F.~Cerutti, M.~P.~W. Chin, A.~Fassò, A.~Ferrari, P.~G. Ortega
  et~al., \emph{{The FLUKA Code: Developments and Challenges for High Energy
  and Medical Applications}},
  \href{https://doi.org/10.1016/j.nds.2014.07.049}{\emph{Nucl. Data Sheets}
  {\bfseries 120} (2014) 211--214}.

\bibitem{NUNDIS}
G.~Battistoni, A.~Ferrari, M.~Lantz, P.~R. Sala and G.~I. Smirnov, \emph{A
  neutrino-nucleon interaction generator for the fluka monte carlo code},  in
  \emph{CERN-Proceedings-2010-001}, pp.~387--394, 2010.

\bibitem{Povinec:2018}
P.~P. Povinec, \emph{{New ultra-sensitive radioanalytical technologies for new
  science}}, \href{https://doi.org/10.1007/s10967-018-5787-3}{\emph{J.
  Radioanal. Nucl. Chem.} {\bfseries 316} (2018) 893--931}.

\bibitem{Povinec:2018wgd}
P.~P. Povinec et~al., \emph{{Ultra-sensitive radioanalytical technologies for
  underground physics experiments}},
  \href{https://doi.org/10.1007/s10967-018-6105-9}{\emph{J. Radioanal. Nucl.
  Chem.} {\bfseries 318} (2018) 677--684}.

\bibitem{Edwards:2017mnf}
T.~D.~P. Edwards and C.~Weniger, \emph{{A Fresh Approach to Forecasting in
  Astroparticle Physics and Dark Matter Searches}},
  \href{https://doi.org/10.1088/1475-7516/2018/02/021}{\emph{JCAP} {\bfseries
  1802} (2018) 021}, [\href{https://arxiv.org/abs/1704.05458}{{\ttfamily
  1704.05458}}].

\bibitem{Edwards:2017kqw}
T.~D.~P. Edwards and C.~Weniger, \emph{{swordfish: Efficient Forecasting of New
  Physics Searches without Monte Carlo}},
  \href{https://arxiv.org/abs/1712.05401}{{\ttfamily 1712.05401}}.

\bibitem{swordfish}
\url{https://github.com/cweniger/swordfish}.

\bibitem{Cowan:2010js}
G.~Cowan, K.~Cranmer, E.~Gross and O.~Vitells, \emph{{Asymptotic formulae for
  likelihood-based tests of new physics}},
  \href{https://doi.org/10.1140/epjc/s10052-011-1554-0}{\emph{Eur. Phys. J. C}
  {\bfseries 71} (2011) 1554},
  [\href{https://arxiv.org/abs/1007.1727}{{\ttfamily 1007.1727}}].

\bibitem{Billard:2012}
J.~{Billard}, F.~{Mayet} and D.~{Santos}, \emph{{Assessing the discovery
  potential of directional detection of dark matter}},
  \href{https://doi.org/10.1103/PhysRevD.85.035006}{\emph{Phys. Rev. D}
  {\bfseries 85} (2012) 035006},
  [\href{https://arxiv.org/abs/1110.6079}{{\ttfamily 1110.6079}}].

\bibitem{Conrad:2014nna}
J.~Conrad, \emph{{Statistical Issues in Astrophysical Searches for Particle
  Dark Matter}},
  \href{https://doi.org/10.1016/j.astropartphys.2014.09.003}{\emph{Astropart.
  Phys.} {\bfseries 62} (2015) 165--177},
  [\href{https://arxiv.org/abs/1407.6617}{{\ttfamily 1407.6617}}].

\bibitem{Wilks:1938dza}
S.~S. Wilks, \emph{{The Large-Sample Distribution of the Likelihood Ratio for
  Testing Composite Hypotheses}},
  \href{https://doi.org/10.1214/aoms/1177732360}{\emph{Annals Math. Statist.}
  {\bfseries 9} (1938) 60--62}.

\bibitem{Agnes:2018ves}
{\scshape DarkSide} collaboration, P.~Agnes et~al., \emph{{Low-Mass Dark Matter
  Search with the DarkSide-50 Experiment}},
  \href{https://doi.org/10.1103/PhysRevLett.121.081307}{\emph{Phys. Rev. Lett.}
  {\bfseries 121} (2018) 081307},
  [\href{https://arxiv.org/abs/1802.06994}{{\ttfamily 1802.06994}}].

\bibitem{Ruppin:2014bra}
F.~Ruppin, J.~Billard, E.~Figueroa-Feliciano and L.~Strigari,
  \emph{{Complementarity of dark matter detectors in light of the neutrino
  background}}, \href{https://doi.org/10.1103/PhysRevD.90.083510}{\emph{Phys.
  Rev.} {\bfseries D90} (2014) 083510},
  [\href{https://arxiv.org/abs/1408.3581}{{\ttfamily 1408.3581}}].

\end{thebibliography}\endgroup

\end{document}